\documentclass[journal,10pt]{IEEEtran}

\IEEEoverridecommandlockouts
\usepackage{cite}
\usepackage{caption}
\usepackage{amsmath,amssymb,amsfonts}
\usepackage{algorithmic}
\usepackage{graphicx}
\usepackage{tabularx}
\usepackage{textcomp}
\usepackage{xcolor}
\usepackage{stfloats}
\usepackage{multirow}
\usepackage{amsmath}
\usepackage{booktabs}
\usepackage{array}
\usepackage{subfigure}
\usepackage{makecell}
\usepackage{array}
\usepackage{booktabs}
\usepackage{overpic}
\def\BibTeX{{\rm B\kern-.05em{\sc i\kern-.025em b}\kern-.08em
		T\kern-.1667em\lower.7ex\hbox{E}\kern-.125emX}}
\begin{document}
\title{A Novel 3D Non-Stationary Channel Model for 6G Indoor Visible Light Communication Systems}
\author{Xiuming Zhu, Cheng-Xiang Wang,~\IEEEmembership{Fellow,~IEEE}, Jie Huang,~\IEEEmembership{Member,~IEEE}, \\ Ming Chen,~\IEEEmembership{Member,~IEEE}, and Harald Haas,~\IEEEmembership{Fellow,~IEEE}
\thanks{This work was supported by the National Key R\&D Program of China under Grant 2018YFB1801101, the National Natural Science Foundation of China (NSFC) under Grants 61960206006 and 61901109, the Frontiers Science Center for Mobile Information Communication and Security, the High Level Innovation and Entrepreneurial Research Team Program in Jiangsu, the High Level Innovation and Entrepreneurial Talent Introduction Program in Jiangsu, the High Level Innovation and Entrepreneurial Doctor Introduction Program in Jiangsu under Grant JSSCBS20210082, the Fundamental Research Funds for the Central Universities under Grant 2242022R10067, the Research Fund of National Mobile Communications Research Laboratory, Southeast University under Grant 2021B02, the EU H2020 RISE TESTBED2 project under Grant 872172, the Royal Society Wolfson Foundation for a Research Merit Award, and EPSRC for an Established Career Fellowship under Grant EP/R007101/1.}
\thanks{X. Zhu, C.-X. Wang (corresponding author), J. Huang, and M. Chen are with the National Mobile Communications Research Laboratory,
School of Information Science and Engineering, Southeast University, Nanjing, 210096, China, and also with the Purple Mountain
Laboratories, Nanjing, 211111, China (email: \{xm\_zhu, chxwang, j\_huang, chenming\}@seu.edu.cn). 

H. Haas is with the LiFi Research and Development Center, Department Electronic and Electrical Engineering, the University of Strathclyde, Glasgow G1 1XQ, U.K. (e-mail: harald.haas@strath.ac.uk).}
}

\markboth{IEEE Transactions on Wireless Communications, vol.~xx, no.~xx, Month~2022}%
{X. Zhu \MakeLowercase{\textit{et al.}}: A Novel 3D Non-Stationary Channel Model for 6G Indoor Visible Light Communication Systems.}

\maketitle

\begin{abstract}
The visible light communication (VLC) technology has attracted much attention in the research of the sixth generation (6G) communication systems. 
In this paper, a novel three dimensional (3D) space-time-frequency non-stationary geometry-based stochastic model (GBSM) is proposed for indoor VLC channels. \textcolor{black}{The proposed VLC GBSM can capture unique indoor VLC channel characteristics such as the space-time-frequency non-stationarity caused by large light-emitting diode (LED) arrays in indoor scenarios, long travelling paths, and large bandwidths of visible light waves, respectively. In addition, the proposed model can support special radiation patterns of LEDs, 3D translational and rotational motions of the optical receiver (Rx), and can be applied to angle diversity receivers (ADRs).} Key channel properties are simulated and analyzed, including the \textcolor{black}{space-time-frequency correlation function (STFCF)}, received power, root mean square (RMS) delay spread, and path loss (PL). Simulation results verify the \textcolor{black}{space-time-frequency} non-stationarity in indoor VLC channels. \textcolor{black}{In addition, the influence of light source radiation patterns, receiver rotations, and ADRs on channel characteristics have been investigated.}
Finally, the accuracy and practicality of the proposed model are validated by comparing the simulation result of channel 3dB bandwidth with the existing measurement data. The proposed channel model will play a supporting role in the design of future 6G VLC systems. 
\end{abstract}


\section{Introduction}
\textcolor{black}{Facing the demand for huge traffic, huge connections, and the need to solve the problem of radio frequency (RF) saturation, the sixth generation (6G) wireless communication systems will fully uitilize all spectrum resources including sub-6 GHz, millimeter wave, terahertz, and optical wireless bands~\cite{wang2020_6g,You2021}. The visible light communication (VLC) has attracted ever-growing attention as a novel technology to explore the visible light optical band to provide higher communication data rate.} In VLC systems, intensity modulation with direct detection (IM/DD) is the most commonly used modulation/demodulation technique\cite{Dimitrov2015}. The information is modulated to the intensity of light emitted from the light-emitting diode (LED) illuminator and demodulated by the photodiode (PD). Up to now, VLC has been studied in multiple application scenarios, e.g., indoor\cite{Karunatilaka2015}, outdoor vehicle-to-vehicle (V2V)\cite{Memedi2021}, underground\cite{Wu2016}, underwater~\cite{Zeng2017, Chen2020}, reconfigurable intelligent surfaces (RIS) aided scenarios~\cite{Abdelhady2021,R_Ndjiongue2021}, etc. In addition, VLC is also used in areas such as indoor positioning\cite{Sheikh2021,Zhuang2018}. Since most of the human activities occur indoors, the indoor scenario is a promising VLC application scenario. Because channel models are essential for the design and evaluation of communication systems\cite{Wang2018_survey}, accurate VLC channel models with low complexity and good pervasiveness are indispensable.

Propagation characteristics of VLC channels were studied in the literature. It was reviewed in \cite{Al-Kinani2018_survey} that the VLC channels do not suffer from \textcolor{black}{small-scale} fading and have negligible Doppler effect. In real VLC systems, different practical LED sources with special radiant patterns are used\cite{Moreno2008}. The wideband nature of visible light waves were studied in \cite{Lee2011,Miramirkhani2020}. The power of incoherent light emitted from a white LED illuminator is distributed over the visible light spectrum, i.e., $380\ {\rm nm}$\ --\ $780\ {\rm nm}$. Therefore, \textcolor{black}{the frequency non-stationarity caused by large bandwidths} needs to be considered in the channel model. In indoor scenarios, the lighting system is usually a large LED array consisting of multiple LED lamps which can be used as multiple optical transmitters (Txs) \cite{Uday2021,Li2021}. \textcolor{black}{To address the co-channel interference, a generalized angle diversity receiver (ADR) structure is proposed in \cite{Chen_ADR_2018} and used at the receiver (Rx) side to reduce the signal to interference plus noise ratio (SINR).} The Rx orientation and its impact on indoor VLC channels were measured and studied in \cite{Soltani2019}. The upper and lower bounds of VLC channel capacity were analyzed in \cite{Wang2013,Jia2020}. In addition, it was reported in \cite{Karunatilaka2015} that reflections of optical signals on indoor materials are mainly diffusive. 

So far, many general standardized channel models for the fifth generation (5G) and beyond 5G (B5G) communication systems have been proposed, e.g., 3GPP TR 38.901 channel model\cite{3GPP}, QuaDRiGa channel model\cite{QuaDRiGa}, IMT-2020 channel model\cite{IMT}, B5GCM\cite{Bian2021}, etc. However, none of these standardized channel models support characterizing VLC channels. In the past few years, a number of VLC channel models were proposed for indoor VLC systems and can be categorized as deterministic and stochastic channel models. In \cite{Lee2011}, an extended indoor VLC channel model based on the classical recursive infrared (IR) channel model \cite{Barry1993} was proposed. The extended recursive model considered the wavelength-dependent white LED radiant power spectral density (PSD) and spectral reflectance of different materials.  In \cite{Ding2016}, influences of three kinds of source radiation patterns on optical wireless channels were investigated based on the IR recursive model\cite{Barry1993}. In \cite{Miramirkhani2015,Miramirkhani2017}, ray tracing channel models based on Zemax$^\circledR$\cite{Zemax} were proposed to model and analyze the VLC channels in indoor scenarios. Both recursive model and ray tracing model are typical accurate channel models, but they are deterministic and lack of pervasiveness. The computational complexity of this kind of channel models will increase greatly when simulating time-varying channels. On the contrary, geometry-based stochastic models (GBSMs) strike a good balance between accuracy, complexity, and pervasiveness, and are widely used in the modeling of wireless communication channels\cite{He2022, Wang2021_THz}. In GBSMs, the interaction between the transmitted signal and the communication environment is abstracted by effective scatterers which are described by geometrical relationships. In \cite{Al-Kinani2016, Al-Kinani2016_2}, regular shaped GBSMs (RS-GBSMs) were proposed for indoor VLC channels. Some statistical properties like root mean square (RMS) delay spread, Ricean factor, and temporal autocorrelation function (ACF) were investigated. \textcolor{black}{However, these VLC RS-GBSMs are all purely geometric and based on ideal assumptions that scatterers are distributed on two dimensional (2D) regular shaped circles and ellipsoids, which are far away from the real communication scenarios.} What's more, all these GBSMs do not consider arbitrary orientations of Rx and are limited to characterize VLC channels with the ideal Lambertian radiation pattern. The wideband nature of visible light waves emitted by white LEDs and non-stationarity of indoor VLC channels are also neglected. 

To the best of the authors’ knowledge, there is no standardized channel model or a comprehensive model which can characterize all the properties of indoor VLC channels. The GBSM is a good candidate for the 6G standardized channel model due to its good tradeoff between accuracy, complexity, and pervasiveness. \textcolor{black}{A lot of standardized channel models including SCM, WINNER, and IMT-Advanced are semi-GBSMs that incorporate user-defined environment, network layout, and antenna array parameters, as well as spatial cluster and scatterer distributions. Compared to pure-GBSMs, these semi-GBSMs are more accurate and scalable. However, existing VLC GBSMs are all purely geometric.} What' more, existing VLC GBSMs do not support three dimensional (3D) translational and rotational motions, arbitrary LED radiation patterns, \textcolor{black}{space-time-frequency} non-stationarity, \textcolor{black}{and limited to the single PD case.} To fill the research gap, a novel GBSM is proposed for indoor VLC systems in this paper. The major contributions and novelties of this work can be summarized as follows.
\begin{enumerate}
	\item {A novel 3D space-time non-stationary GBSM is proposed for indoor VLC channels. The proposed model extends the existing 2D RS-GBSM into 3D semi-GBSM which is more accurate. \textcolor{black}{The novel GBSM can support  special radiation patterns of LEDs, 3D motion speeds of clusters and Rx, arbitrary orientations or rotations of the optical Rx, and can be applied to ADRs.}}
	\item {\textcolor{black}{The non-stationarities of indoor VLC channels in the spatial, time, and frequency domain caused by the large LED array, continuous movement of the optical Rx, and large bandwidths are considered and modeled by taking the cluster evolution in the space domain, space- and time-varying channel parameters, and the wavelength-dependency of visible light waves into consideration.}}
	\item {Based on the proposed model, key statistical properties such as the \textcolor{black}{space-time-frequency correlation function (STFCF)}, received power, RMS delay spread, and path loss (PL) are studied and analyzed. Influences of key model parameters on channel statistical properties are investigated. \textcolor{black}{The effects of special LED radiation patterns, Rx rotations, and ADRs on the channel are explored and discussed.} 
Finally, the simulation result of channel 3dB bandwidth of the proposed model can fit well with existing measurement data, showing the accuracy and practicality of the model.}
\end{enumerate}

The remainder of this paper is organized as follows. In Section~\uppercase\expandafter{\romannumeral2}, a novel \textcolor{black}{space-time-frequency} non-stationary GBSM for indoor \textcolor{black}{MIMO} VLC systems is described in detail. The indoor VLC system model, channel impulse response (CIR), and the generation of channel coefficient are introduced. In Section~\uppercase\expandafter{\romannumeral3}, key statistical properties of the model are investigated. Section~\uppercase\expandafter{\romannumeral4} presents simulation results and discussions. Finally, conclusions are drawn in Section~\uppercase\expandafter{\romannumeral5}.

\section{A Novel 3D Indoor VLC GBSM}
\subsection{System Model}
\begin{figure}[t]
	\centerline{\includegraphics[width=0.4\textwidth]{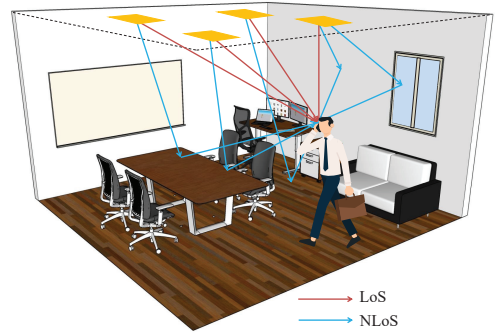}}
	\caption{A typical indoor VLC propagation scenario.}
	\label{fig_1}
\end{figure}

Compared with traditional RF wireless indoor communication systems, the Tx and Rx of VLC systems are no longer RF antennas but \textcolor{black}{LED lamps and PDs}, respectively. Fig.~{\ref{fig_1}} shows a typical indoor VLC propagation scenario with the line-of-sight (LoS) and non-LoS (NLoS) paths. The propagation scenario is a meeting room designed in SketchUp$^\circledR$ \cite{SU_web}. In indoor scenarios, the lighting system usually consists of multiple LED lamps which support the illumination for the room. By means of the VLC technology, these LED lamps can be used for both illumination and communication. \textcolor{black}{At the Rx side, a single PD or an ADR\cite{Chen_ADR_2018,Zeng_WCNC_2019} consisting of multiple PDs can be set on the mobile phone.} In our work, we focus on a typical indoor  VLC scenario illustrated in Fig.~{\ref{fig_1}}. In reality, a LED lamp consists of many LED chips. It is assumed that each LED lamp consisting of multiple chips has a unified radiation pattern. When the user moves or rotates the mobile phone, the \textcolor{black}{Rx} on the mobile phone will move and rotate simultaneously. After the optical signal carrying information is emitted from the LED, it may reach the Rx directly or through some reflections. It is worth noting that optical signals will reflect diffusely on rough objects (e.g., plaster wall and wood floor) while specular reflections can occur on some smooth objects (e.g., glass window). However, it has been reported in \cite{Karunatilaka2015,Lee2011} that reflections of optical signals on most of indoor objects are typically diffuse. In our work, it is assumed that optical signals experience purely diffusive reflections on indoor materials.

In contrast to conventional RF-based communication systems, another feature of VLC systems is that IM/DD is the most common modulation/demodulation technique for VLC \cite{Dimitrov2015}. This is a necessity because light waves emitted by LEDs are incoherent \cite{Cheng2018}. Since the intensity of an optical signal is actually related to the optical power of the signal, the transmitted signal $x(t)$ is the optical power signal\cite{Ghassemlooy2013}. For a time-varying IM/DD-based VLC system, the relationship between the transmitted optical power signal $x(t)$ and the received signal $y(t)$ can be written as 
\begin{equation}
	y(t)=\int_{-\infty}^{\infty} x(t-\tau) h(t,\tau) d\tau + n(t)
\end{equation}
\textcolor{black}{where $h(t,\tau)$ is the response of the channel at time $t$ to an impulse at time $t-\tau$ characterizing the time-varying optical power loss, $\tau$ denotes the delay, and $n(t)$ is the noise.}

In this paper, we focus on the modeling of $h(t,\tau)$ of indoor VLC systems described above. Like several standardized channel models for 5G communication systems, we consider the semi-GBSM modeling approach. In the next section, we will give a more detailed description of the proposed model. 

\subsection{\textcolor{black}{Description of the GBSM}}
\begin{figure*}[tb]
	\centerline{\includegraphics[width=0.95\textwidth]{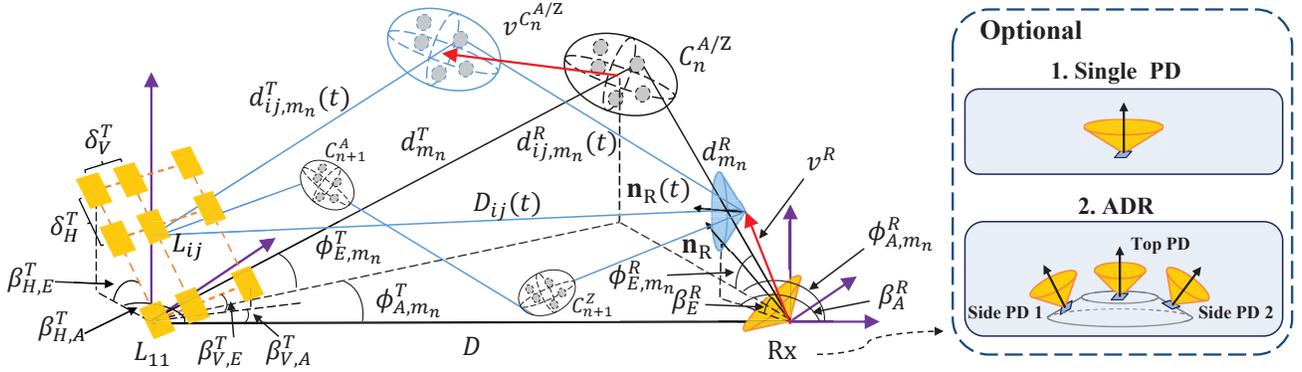}}
	\caption{\textcolor{black}{A 3D non-stationary GBSM for indoor VLC systems.}}
	\label{fig_2}
\end{figure*}
The proposed novel 3D GBSM is presented in Fig.~{\ref{fig_2}} where a uniform planar LED array is set at the Tx side and \textcolor{black}{the Rx can be a single PD or an ADR consisting of multiple PDs optionally. We consider a generalized ADR proposed in \cite{Chen_ADR_2018} which is particularly suitable for handheld terminals \cite{Zeng_WCNC_2019}. The generalized ADR is composed of totally $N_{\rm PD}$ PDs, including a PD at the top and $N_{\rm PD}-1$ inclined side PDs uniformly distributed on a circumference. In our work, we mainly consider the scenario where the ADR is installed on a handheld terminal and the size of the ADR is typically small. Therefore, the tiny distance difference between each PD in the ADR is ignored in the proposed model, and elements in the ADR are assumed to have the same location but different orientations. For clarity, only two side PDs are illustrated in the ADR case in Fig.~{\ref{fig_2}}. The elevation angle difference between the normal vector of the top PD and those of side PDs is denoted as $\theta_{\rm PD}$.} There are $M_I$ rows and $M_J$ columns in the LED array where the spacing between LED elements in a row and in a column are $\delta_V^T$ and $\delta_H^T$, respectively. The LED element at the $i$-th row and the $j$-th column is denoted as $L_{ij}$. The azimuth and elevation angles of the row (column) of the LED array are denoted as $\beta_{V(H),A}^T$ and $\beta_{V(H),E}^T$, respectively. \textcolor{black}{At the Rx side, the normal vector of the single PD or the top PD of the ADR at the initial time is denoted as $\mathbf{n}_{\rm R}$ with the azimuth and elevation angles denoting as $\beta_A^R$ and $\beta_E^R$, respectively. Since the \textcolor{black}{Rx} in a VLC system has random orientations, the normal vector of the \textcolor{black}{Rx}  at time $t$ is denoted as $\mathbf{n}_{\rm R}(t)$.} \textcolor{black}{It has been reported in \cite{Miramirkhani2020} that higher-order reflections make negligible contribution to the total CIR. In the proposed model, the single-bounce (SB) and double-bounce (DB) NLoS propagation are considered. For the sake of simplicity, only the $n$-th path representing SB components and the $n+1$-th path representing DB components are illustrated in Fig.~{\ref{fig_2}}. The NLoS propagation is abstracted by the cluster $C_{n}^A$ at the Tx side and the cluster $C_{n}^Z$ at the Rx side. Note that $C_{n}^A$ and $C_{n}^Z$ are the same cluster for SB components. Totally, there are $N_{ij}(t)$ paths between $L_{ij}$ and the Rx at time $t$. Within $C_n^{A/Z}$, there are $M_n$ scatterers.} In indoor scenarios, the LED array acts as both an illuminator and a communication device, so the Tx is static while clusters and the Rx can be moving. The moving velocities of clusters (the Rx) are assumed to be constant, \textcolor{black}{with speed $v^{{C_n^{A/Z}}(R)}$, azimuth travel angle $\alpha_A^{{C_n^{A/Z}}(R)}$, and elevation travel angle $\alpha_E^{{C_n^{A/Z}}(R)}$.} The rotation azimuth velocity and elevation velocity of the \textcolor{black}{Rx} are denoted as $\omega_A^R$ and $\omega_E^R$, respectively. The azimuth angle of departure (AAoD) and elevation angle of departure (EAoD) of the $m$-th ray in $C_n^A$ transmitted from $L_{11}$ are denoted by $\phi_{A,m_n}^T$ and $\phi_{E,m_n}^T$, respectively. The azimuth angle of arrival (AAoA) and elevation angle of arrival (EAoA) of the $m$-th ray in the $C_n^Z$ impinging on the \textcolor{black}{Rx} are denoted by $\phi_{A,m_n}^R$ and $\phi_{E,m_n}^R$, respectively. \textcolor{black}{Denoting the $m$-th scatterer in $C_n^{A/Z}$ as $S_{m_n}^{A/Z}$, then the transmission distances $L_{11}$-$S_{m_n}^{A}$, $S_{m_n}^{A}$-$S_{m_n}^{Z}$ (for DB case), and $S_{m_n}^{Z}$-Rx are denoted by $d_{m_n}^T$, $d_{m_n}^S$, and $d_{m_n}^R$, respectively.} Similarly, the angle parameters and transmission distance of the LoS path from $L_{11}$ to \textcolor{black}{the Rx} are denoted by $\phi_{A(E),L}^{T(R)}$ and $D$, respectively. Note that the angle and distance parameters defined above are the values at the initial time instance, all these parameters are modeled as time-varying in the proposed model. Key parameters of the proposed model are summarized in Table {\ref{table_1}}.

\renewcommand{\arraystretch}{1.3}
\begin{table*}[tb]
	\centering
	\caption{\textcolor{black}{Definition of key parameters of the proposed model.}}
	\begin{tabular}{|l|l|}

		\hline
		\textbf{Parameter}& \textbf{Definition} \\ \hline
		$L_{ij}$ & The LED element at the $i$-th row and the $j$-th column \\ \hline
		$C_n^{A/Z}, S_{m_n}^{A/Z}$ & The $n$-th cluster at the Tx/Rx side and the $m$-th scatterer in the $n$-th cluster \\ \hline
		$\delta_V^T$, $\delta_H^T$ & Spacing between LED elements in a row and in a column, respectively \\ \hline
		$\beta_{V(H),A}^T$, $\beta_{V(H),E}^T$ & Azimuth and elevation angles of the row (column) of the LED array, respectively \\ \hline
		$\beta_A^R$, $\beta_E^R$ & Azimuth and elevation angles of the normal vector of the \textcolor{black}{Rx} at the initial time \\ \hline
		$\beta_A^R(t)$, $\beta_E^R(t)$ & Azimuth and elevation angles of the normal vector of the \textcolor{black}{Rx} at time $t$ \\ \hline
		$v^{{C_n^{A/Z}}(R)}$, $\alpha_A^{{C_n^{A/Z}}(R)}$, $\alpha_E^{{C_n^{A/Z}}(R)}$ & Speeds, travel azimuth angles, and travel elevation angles of $C_n^{A/Z}$ (Rx), respectively \\ \hline
		$\omega_A^R$, $\omega_E^R$ & Rotation azimuth velocity and elevation velocity of the \textcolor{black}{Rx}, respectively \\ \hline
		$\phi_{A,L}^{T(R)}$, $\phi_{E,L}^{T(R)}$ & AAoD (AAoA) and EAoD (EAoA) of the LoS path from $L_{11}$ to the Rx at the initial time \\ \hline
		$\phi_{ij,A,L}^{T(R)}(t)$, $\phi_{ij,E,L}^{T(R)}(t)$ & AAoD (AAoA) and EAoD (EAoA) of the LoS path from $L_{ij}$ to the Rx at time $t$ \\ \hline
		$\phi_{A,m_n}^T$, $\phi_{E,m_n}^T$ & AAoD and EAoD of the $m$-th ray in $C_n^A$ transmitted from $L_{11}$ at the initial time \\ \hline
		$\phi_{A,m_n}^R$, $\phi_{E,m_n}^R$ & AAoA and EAoA of the $m$-th ray in $C_n^Z$ impinging on the Rx at the initial time \\ \hline
		$\phi_{ij,A,m_n}^T(t)$, $\phi_{ij,E,m_n}^T(t)$ & AAoD and EAoD of the $m$-th ray in $C_n^A$ transmitted from $L_{ij}$ at time $t$ \\ \hline
		$\phi_{A,m_n}^R(t)$, $\phi_{E,m_n}^R(t)$ & AAoA and EAoA of the $m$-th ray in $C_n^Z$ impinging on the Rx at time $t$ \\ \hline
		$d_{m_{n}}^{T}$, $d_{m_{n}}^{S}$, $d_{m_{n}}^{R}$ & Transmission distances $L_{11}$-$S_{m_n}^A$, $S_{m_n}^A$-$S_{m_n}^Z$, and $S_{m_n}^Z$-Rx at the initial time \\ \hline
		$d_{ij,m_{n}}^{T}(t)$, $d_{m_{n}}^{S}(t)$, $d_{m_{n}}^{R}(t)$ & Transmission distances $L_{ij}$-$S_{m_n}^A$, $S_{m_n}^A$-$S_{m_n}^Z$, and $S_{m_n}^Z$-Rx at time $t$ \\ \hline
		$D$, $D_{ij}(t)$ & Transmission distance $L_{11}$-Rx at the initial time and distance $L_{ij}$-Rx at time $t$ \\ \hline
		$N_{ij}(t)$, $M_n$ & Number of paths between $L_{ij}$ and the Rx at time $t$ and number of scatterers within $C_n^{A/Z}$ \\ \hline
	\end{tabular}
	\label{table_1}
\end{table*}

\subsection{CIR}
\textcolor{black}{Compared with RF channels, one of the most notable features of VLC channels is that there is no small-scale fading in VLC channels undergoing multipath propagations\cite{Al-Kinani2018_survey, Dimitrov2015}. On the one hand, the wavelength of visible light signals is extremely short, and the size of the receiver is usually millions of square wavelengths. Therefore, the fast fading of signals over several wavelengths will not occur. On the other hand, the superposition of real-valued optical power signals after multipath propagation will not cause the constructive and destructive fast fading, but will lead to the slow fading in the form of shadowing\cite{Dimitrov2015}. 
In indoor VLC channels, the large-scale path loss $PL$, shadowing $SH$, and blockage effect $BL$ need to be taken into consideration. \textcolor{black}{Note that $BL$ is typically determined and modeled by the blockage probability as in~\cite{Chen2021}}.}


%

\textcolor{black}{Considering multiple LEDs as the Tx and an ADR consisting of multiple PDs as the Rx, the complete channel matrix is represented by $\mathbf{H}=\left[h_{ij,p,\lambda_T}(t,\tau)\right]_{M_I \times M_J \times N_{\rm PD}}$ where $h_{ij,p,\lambda_T}(t,\tau)$ is the CIR of the sub-channel between $L_{ij}$ and the $p$-th PD and $\lambda_T$ denotes the wavelength range of the light source. The sub-channel impulse response is the superposition of the LoS and NLoS components. Since only NLoS components are related to the color of the light source, the LoS and NLoS components can be written as
\begin{equation}
	h_{ij,p,\lambda_T}(t,\tau)=h_{ij,p}^{\rm L}(t,\tau)+h_{ij,p,\lambda_T}^{\rm N}(t,\tau).
\end{equation}
Due to the fact that transmitted signals in VLC channels are real-valued optical power signals, the CIR equations of the LoS and NLoS components of VLC channels can be expressed as
\begin{equation}\label{CIR_L_new}
	h_{ij,p}^{\rm L}(t,\tau)=P_{ij,p}^{\rm L}(t)\cdot \delta \left( \tau-\tau_{ij,p}^{\rm L}(t) \right)
\end{equation}
\begin{equation}\label{CIR_N_new}
	h_{ij,p,\lambda_T}^{\rm N}(t,\tau)=\sum_{n=1}^{N_{ij}(t)}\sum_{m=1}^{M_n}P_{ij,p,\lambda_T,m_n}^{\rm N}(t) \cdot \delta \left( \tau-\tau_{ij,p,m_n}^{\rm N}(t) \right).
\end{equation}
Here, $P_{ij,p}^{\rm L}(t)$ ($P_{ij,p,\lambda_T,m_n}^{\rm N}(t)$) and $\tau_{ij,p}^{\rm L}(t)$ ($\tau_{ij,p,m_n}^{\rm N}(t)$) are the actual power loss and propagation delay of the LoS path (the $m$-th ray in the $n$-th path) from $L_{ij}$ to the $p$-th PD, respectively.}

\begin{figure*}[t]
	\centering
	\subfigure[]{{\includegraphics[width=0.38\textwidth]{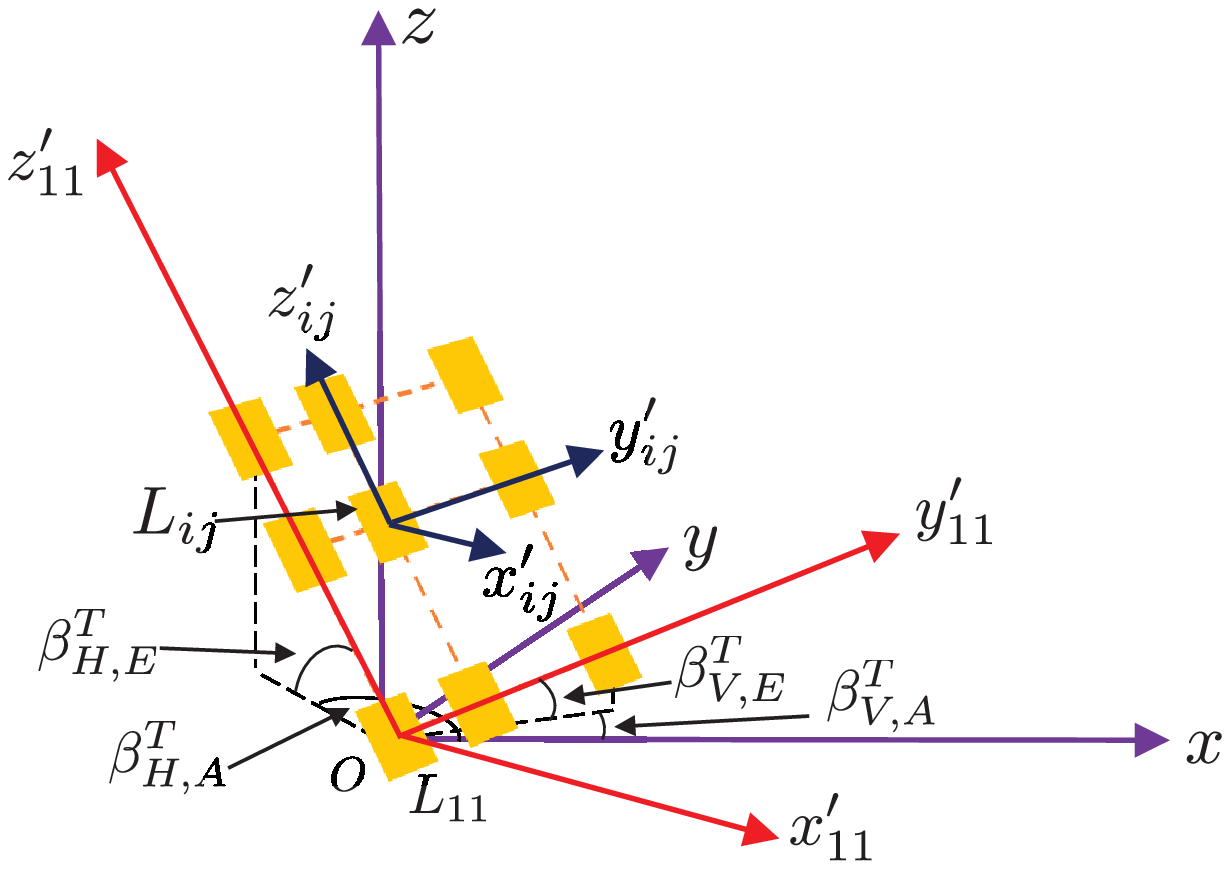}}\label{fig_3a}}
	\subfigure[]{{\includegraphics[width=0.26\textwidth]{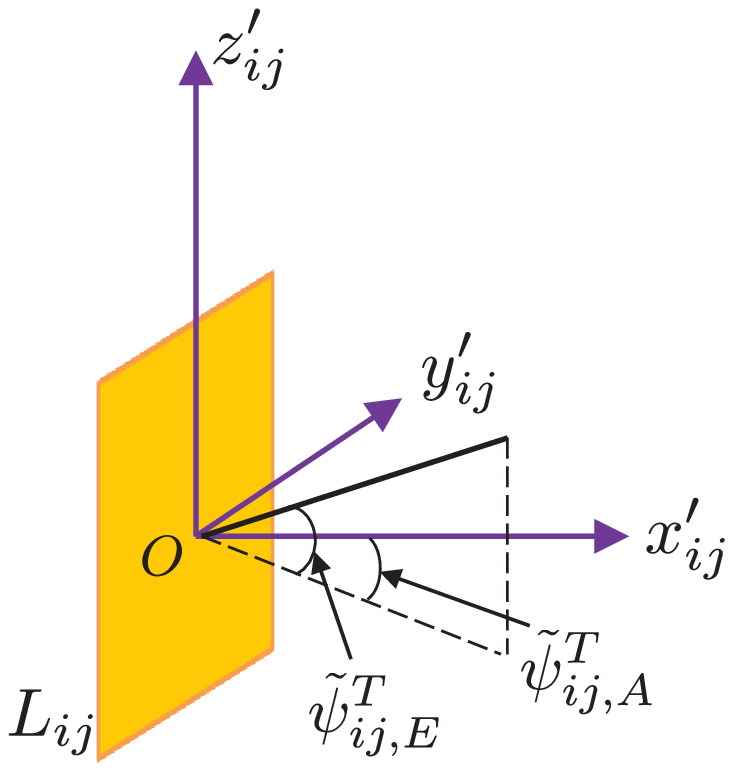}}\label{fig_3b}}
	\caption{The illustration of a) the GCS, LCS of $L_{11}$, and LCS of $L_{ij}$, b) angles in the LCS of $L_{ij}$.}
	\label{fig_3}
\end{figure*}

For the LoS component, the actual ray power is calculated according to the LED radiation pattern as 
\begin{align}\label{power_L}
	P_{ij,p}^{\rm L}(t)&=F_{ij}\left( \tilde{\psi}_{ij,E,{\rm L}}^{T}(t),\tilde{\psi}_{ij,A,{\rm L}}^{T}(t) \right) \cdot \frac{A_R\cos \left( \psi_{ij,p,{\rm L}}^R(t) \right)}{\left(D_{ij}(t)\right)^2} \notag \\
	&\cdot G\left(\psi_{ij,p,{\rm L}}^{R}(t)\right) T\left(\psi_{ij,p,{\rm L}}^{R}(t)\right) V\left(\psi_{ij,p,{\rm L}}^{R}(t)\right).
\end{align}
Here, $F_{ij}(\tilde{\psi}_{ij,E}^T,\tilde{\psi}_{ij,A}^T)$ is the normalized radiation pattern of $L_{ij}$ which is defined as the optical power per unit solid angle (W/sr) emitted from the source with the total power of $1\ {\rm W}$ in a given direction with elevation and azimuth angles of $\tilde{\psi}_{ij,E}^T$ and $\tilde{\psi}_{ij,A}^T$, respectively. In~(\ref{power_L}), $A_R$ is the area of the PD, $\tilde{\psi}_{ij,E,{\rm L}}^{T}(t)$ and $\tilde{\psi}_{ij,A,{\rm L}}^{T}(t)$ are EAoD and AAoD of LoS path from $L_{ij}$ to the optical Rx in the local coordinate system (LCS) of $L_{ij}$ element, \textcolor{black}{$\psi_{ij,p,{\rm L}}^R(t)$} is the angle between the LoS path and the normal of the \textcolor{black}{$p$-th PD}. $G(\psi^R)$, $T(\psi^R)$, and $V(\psi^R)$ are the optical gain of lens, optical filter gain, and visible function at Rx side, respectively. Note that $\tilde{\psi}_{ij,A}^T$ and $\tilde{\psi}_{ij,E}^T$ in this paper represent angles in the LCS of each LED element. They need to be transformed from the global coordinate system (GCS). In the LCS of $L_{ij}$, the $x_{ij}^{\prime}$ axis is defined as the normal of $L_{ij}$, and the $y_{ij}^{\prime}oz_{ij}^{\prime}$ plane is defined as the plane where the LED element is located. \textcolor{black}{Fig.~{\ref{fig_3a}} gives an illustration of the GCS ($x,y,z$), the LCS ($x_{11}^{\prime},y_{11}^{\prime},z_{11}^{\prime}$) of $L_{11}$, and the LCS ($x_{ij}^{\prime},y_{ij}^{\prime},z_{ij}^{\prime}$) of $L_{ij}$. The LCS of $L_{ij}$ and corresponding angles are shown in Fig.~{\ref{fig_3b}}.} Transformation of angles from the GCS to the LCS will be discussed in detail in the following sections.

%

In previous studies, existing VLC channel models usually only consider the ideal Lambertian radiation pattern of LED lamp. However, practical commercial LEDs have their own specific radiation patterns. In the proposed model, the LED elements can have any radiation pattern. Some mathematical radiation pattern models for common LEDs can be found in \cite{Moreno2008}. In other cases, the luminous intensity in unit of lumen per unit solid angle (lm/sr) of a practical LED $L_{ij}(\tilde{\psi}_{ij,E}^T,\tilde{\psi}_{ij,A}^T)$ can be measured and the actual radiation pattern can be obtained by 
\begin{equation}
	\tilde{F}_{ij}(\tilde{\psi}_{ij,E}^T,\tilde{\psi}_{ij,A}^T)=\frac{L_{ij}(\tilde{\psi}_{ij,E}^T,\tilde{\psi}_{ij,A}^T)}{\rm LER}
\end{equation}
where LER is the luminous efficacy of radiation in unit of lumen per Watt (lm/W). Then the actual radiation pattern needs to be normalized before it is applied to (\ref{power_L}). For the ideal Lambertian radiation pattern, it can be expressed as \cite{Gfeller1979}
\begin{equation}\label{Lambertian}
	F_{ij}\left(\tilde{\psi}_{ij, E}^{T}, \tilde{\psi}_{ij, A}^{T}\right)=\frac{\alpha+1}{2 \pi} \cos ^{\alpha}(\tilde{\psi}_{ij,E}^{T}) \cos ^{\alpha} (\tilde{\psi}_{ij,A}^{T})
\end{equation}
where $\alpha$ stands for the Lambertian radiation mode number. Note that the coordinate system in this paper is different from that in \cite{Gfeller1979}, so the equation of the Lambertian radiation pattern is rewritten as (\ref{Lambertian}) according to the angle transformation.

In VLC systems, the non-imaging concentrator with gain $G(\psi^R)$ and optical filter with gain $T(\psi^R)$ are usually used to enhance the system performance. The optical gain of concentrator lens $G(\psi^R)$ can be calculated as \cite{Ghassemlooy2013}
\begin{equation}
	G(\psi^R)=\left\{\begin{array}{ll}
		\frac{n_{\rm ind}^2}{\sin^2(\psi^R)}, & 0 \leq \psi^R \leq \Psi_{\mathrm{FoV}} \\
		0, & \psi^R>\Psi_{\mathrm{FoV}}
		\end{array}\right.
\end{equation}
where $n_{\rm ind}$ denotes the lens refractive index and $\Psi_{\mathrm{FoV}}$ is the field of view (FoV) of the PD. If the VLC system is not equipped with a concentrator nor an optical filter, then corresponding optical gains are set as 1, i.e., $G(\psi^R)=T(\psi^R)=1$. At last, the visible function is defined making sure that only rays within the PD's FoV can be received, i.e.,
\begin{equation}
	V(\psi^R)=\left\{\begin{array}{ll}
		1, & 0 \leq \psi^R \leq \Psi_{\mathrm{FoV}} \\
		0, & \psi^R>\Psi_{\mathrm{FoV}}.
		\end{array}\right.
\end{equation}

For \textcolor{black}{SB} NLoS components, the CIR is calculated as follows. Firstly, each effective scatterer is considered as a Rx with an effective area $A_{m_n,{\rm eff}}$. Then it is considered as an optical source whose radiation pattern is described by how signals will reflect off it. Consequently, the actual power of each \textcolor{black}{SB} ray is given by
\begin{align}\label{power_N}
	P_{ij,p,\lambda_T,m_n}^{\rm N}(t)&= F_{ij} \left(\tilde{\psi}_{ij,E,m_{n}}^{T}(t),\tilde{\psi}_{ij,A,m_{n}}^{T}(t)\right) \notag \\
	&\times \frac{A_{m_{n}, \mathrm{eff}} \cos \left(\psi_{ij,m_{n}}^{S,T}(t)\right)}{\left(d_{ij,m_{n}}^{T}(t)\right)^{2}} \notag \\
	&\times \Gamma_{ij,\lambda_T,n}R\left(\psi_{m_n}^{S,R}(t)\right) \frac{A_{R} \cos \left(\psi_{p,m_{n}}^{R}(t)\right)}{\left(d_{m_{n}}^{R}(t)\right)^{2}} \notag \\
	&\times G\left(\psi_{p,m_{n}}^{R}(t)\right) T\left(\psi_{p,m_{n}}^{R}(t)\right) V\left(\psi_{p,m_{n}}^{R}(t)\right)
\end{align}
where $\tilde{\psi}_{ij,E,m_{n}}^{T}(t),\tilde{\psi}_{ij,A,m_{n}}^{T}(t)$ are the EAoD and AAoD of the $m$-th ray in \textcolor{black}{$C_n^{A/Z}$} from $L_{ij}$ to Rx in the LCS of $L_{ij}$, 
$\psi_{ij,m_n}^{S,T}(t)$ ($\psi_{m_n}^{S,R}(t)$) is the AoA (AoD) of the wave impinging on (reflecting off) the \textcolor{black}{$S_{m_n}^{A/Z}$}, \textcolor{black}{$\psi_{p,m_{n}}^{R}(t)$} is the angle between the the $m$-th ray in \textcolor{black}{$C_n^{A/Z}$} and the normal of the \textcolor{black}{$p$-th PD}, $\Gamma_{ij,\lambda_T,n}$ denotes the effective reflectance of \textcolor{black}{$C_n^{A/Z}$} considering wavelength-dependent properties \textcolor{black}{of the light source}, $R\left(\psi_{m_n}^{S,R}(t)\right)$ is the reflection model characterizing the reflecting power of each ray from $S_{m_n}$. \textcolor{black}{Theoretically, the number of rays in the cluster is infinite\cite{Liu2020}. Considering the tradeoff between accuracy and complexity, the number of rays in each cluster $M_n$ can be set as 50 or 100 \cite{Wang2021}.}


Since the power of incoherent visible light emitted by a LED illuminator is distributed over corresponding spectral range, wavelength-dependent properties of VLC channels need to be considered. In~\cite{Lee2011}, an effective reflectance parameter considering wavelength-dependent radiant PSDs of white LEDs and wavelength-dependent reflectance parameters of materials over the white light spectrum is introduced. In the proposed model, the parameter is generalized to support VLC systems using LEDs of any color as the Tx and the effective reflectance parameter of \textcolor{black}{$C_n^{A/Z}$} for $L_{ij}$-PDs sub-channels is given by 
\begin{equation}\label{Gamma}
	\Gamma_{ij,\lambda_T,n}=\int_{\lambda_1}^{\lambda_2} \Phi_{ij}(\lambda) \rho_{n}(\lambda) d \lambda
\end{equation}
where $\Phi_{ij}(\lambda)$ is the normalized wavelength-dependent radiant PSD of $L_{ij}$, $\rho_n(\lambda)$ denotes the wavelength-dependent reflectance of \textcolor{black}{$C_n^{A/Z}$, $\lambda_T = [\lambda_1,\lambda_2]$} is the wavelength range of the LED. The wavelength-dependent data for $\Phi_{ij}(\lambda)$ and $\rho_n(\lambda)$ of some common materials in indoor scenarios can be found in \cite{Lee2011,Miramirkhani2020,XLamp}. The diffuse reflections are modeled as \cite{Lee2011} 
\begin{equation}
	R\left(\psi_{m_n}^{S,R}(t)\right)=\frac{1}{\pi}\cos\left(\psi_{m_n}^{S,R}(t)\right).
\end{equation}

\textcolor{black}{For DB NLoS components, the CIR is calculated in a similar way but it contains a part of power loss of the $S_{m_n}^A-S_{m_n}^Z$ link. Due to the limited space, the detailed calculation of the power of each DB ray is not shown here.
}


The delays of the LoS component and NLoS components are determined by
\begin{equation}\label{tau_L}
	\tau_{ij}^{\rm L}(t)=D_{ij}(t)/c_l
\end{equation}
\begin{equation}\label{tau_N}
	\tau_{ij,m_n}^{\rm N}(t)=\begin{cases}
		\left(d_{ij,m_n}^T(t)+d_{m_n}^R(t)\right)/c_l, & \text{SB case} \\
		\left(d_{ij,m_n}^T(t)+d_{m_n}^S(t)+d_{m_n}^R(t)\right)/c_l, & \text{DB case}.
		\end{cases}
\end{equation}
Here, $c_l$ denotes the propagation speed of light.

\begin{figure}[t]
	\centerline{\includegraphics[width=0.52\textwidth]{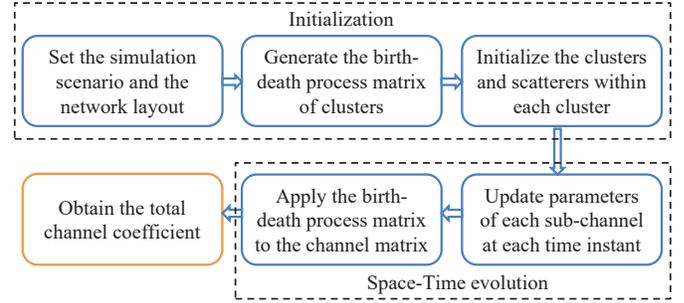}}
	\caption{Flow chart of the channel coefficient generation.}
	\label{fig_4}
\end{figure}

The flow chart of the channel coefficient generation is illustrated in Fig.~{\ref{fig_4}}. 
In the initialization stage, the scenario and layout parameters are preset, then the birth-death process matrix showing the visibility of LED elements to the clusters is generated \textcolor{black}{as will be introduced in Section II. D}. Next, clusters and scatterers are randomly generated and initialized \textcolor{black}{based on assumptions given in Section II. E}. Then, the space- and time-varying channel parameters can be updated according to geometrical relationships \textcolor{black}{which are shown in Section II. F}. Lastly, the birth-death process matrix is applied to the whole channel matrix to set the contributions of unobservable links as zero and the total channel coefficient can be obtained. 


\subsection{Cluster Evolution in the Space Domain}
In indoor VLC systems, multiple LED lamps can be used for communication. 
Although the total number of LED lamps is small, different LED lamps with certain distances apart may observe different clusters, resulting a birth-death behavior over the large uniform planar LED array. Based on the cluster evolution on uniform linear array proposed in \cite{Bian2021}, we consider the cluster evolution in the space domain in two directions, i.e., the horizontal and vertical direction of LED array. The birth-death process has two key parameters, i.e., the cluster birth rate $\lambda_B$ and the cluster death rate $\lambda_D$. For the benchmark LED element, the number of observable clusters at the initial time is determined by $N_{c0}=\lambda_B/ \lambda_D$. For the horizontal direction, the probability of a cluster remaining over LED element spacing $\delta_H^T$ is given as \cite{Bian2021}
\begin{equation}
	P_{H, {\rm remain}}(\delta_H^T)=\exp\left(-\lambda_B \cdot \frac{\delta_H^T\cos(\beta_{H,E}^T)}{D_c^A}\right)
\end{equation}
where $D_c^A$ is the array correlation factor relevant to the specific scenario. Similarly, the probability of a cluster surviving over spacing $\delta_V^T$ in the vertical direction is computed as
\begin{equation}
	P_{V, {\rm remain}}(\delta_V^T)=\exp\left(-\lambda_B \cdot \frac{\delta_V^T\cos(\beta_{V,E}^T)}{D_c^A}\right).
\end{equation}
The number of newly generated clusters is assumed to obey Poisson distribution and the mean value is given as
\begin{equation}
	\mathbb{E}(N_{\rm New,H(V)})=\frac{\lambda_B}{\lambda_D}\left[1-P_{H(V), {\rm remain}}\left(\delta_{H(V)}^T\right)\right].
\end{equation}

In the initialization stage of channel coefficient generation, the birth-death process matrix is generated randomly according to the aforementioned assumptions. It is a 3D matrix sized $M_I\times M_J \times N_{c,{\rm total}}$, showing the visibility of $N_{c,{\rm total}}$ clusters to each LED element. Firstly, the cluster evolution on the first column ($L_{11}-L_{{M_I}1}$) in the horizontal direction is simulated based on $L_{11}$. Then the evolution on each column in the vertical direction is generated based on the first column. Note that the total number of clusters $N_{c,{\rm total}}$ is the sum of $N_{c0}$ and the number of newly generated clusters. In the last step of channel coefficient generation, the obtained birth-death matrix is multiplied to the whole channel matrix. Thus, the contributions of unobservable links are set as zero.

\subsection{Initialization of Clusters and Scatterers}
In the proposed model, clusters and scatterers are randomly generated at the initial time, then channel parameters are updated according to geometrical relationships. In the GCS, the initial location of \textcolor{black}{a cluster} is completely determined by three parameters, i.e., AAoD\textcolor{black}{/AAoA}, EAoD\textcolor{black}{/EAoA}, and distance. The angle parameters are assumed to have wrapped Gaussian distributions. \textcolor{black}{For example, the angles of $C_n^A$} can be obtained as
\begin{equation}
	\phi_{E,n}^T={\rm std}[\phi_{E,n}^{T}]Y_{E,n}^T+\bar{\phi}_{E,n}^T
\end{equation}
\begin{equation}
	\phi_{A,n}^T={\rm std}[\phi_{A,n}^{T}]Y_{A,n}^T+\bar{\phi}_{A,n}^T
\end{equation}
where $Y_{E,n}^T,Y_{A,n}^T \backsim \mathcal{N}(0,1)$, ${\rm std}[\phi_{E,n}^{T}] ({\rm std}[\phi_{A,n}^{T}])$ and $\bar{\phi}_{E,n}^T$ ($\bar{\phi}_{A,n}^T$) are standard deviation and mean value of EAoD (AAoD), respectively. The distance from $L_{11}$ to \textcolor{black}{$C_n^A$} denoted as $d_{n}^T$ is assumed to be a non-negative random variable with an exponential distribution. \textcolor{black}{In the simulation, totally $N_{c,{\rm total}}$ clusters at the Tx side are generated based on above assumptions firstly, then the same process is used to generate $N_{c,{\rm total}} \times (1-\eta_{\rm SB})$ clusters at the Rx side, where $\eta_{\rm SB}$ is the ratio of SB components in NLoS paths.}

For the distribution of effective scatterers within clusters, a general 3D ellipsoid Gaussian scattering distribution \cite{Bian2021} is applied to describe the scatterers. The probability density function of scatterers' coordinates relative to the center point of a cluster $[x',y',z']^{\rm T}$ is given as \cite{Bian2021}
\begin{equation}
	p\left(x^{\prime}, y^{\prime}, z^{\prime}\right)=\frac{\exp \left(-\frac{x^{\prime 2}}{2 \sigma_{D S}^{2}}-\frac{y^{\prime 2}}{2 \sigma_{A S}^{2}}-\frac{z^{\prime 2}}{2 \sigma_{E S}^{2}}\right)}{(2 \pi)^{3 / 2} \sigma_{D S} \sigma_{A S} \sigma_{E S}}
\end{equation}
where $\sigma_{DS}$, $\sigma_{AS}$, and $\sigma_{ES}$ stand for the standard deviations in three directions characterizing the delay spread, angular spread, and elevation spread of the cluster. According to transformation from LCS of the cluster to GCS, the cartesian coordinates of scatterers within a cluster in GCS $[x,y,z]^{\rm T}$ can be obtained as \cite{Bian2021}
\begin{align}\label{coord_s}
	\begin{bmatrix}
		x\\y\\z
	\end{bmatrix} &= \begin{bmatrix}
		\cos \left(\bar{\phi}_{A}\right) & -\sin \left(\bar{\phi}_{A}\right) & 0 \\
		\sin \left(\bar{\phi}_{A}\right) & \cos \left(\bar{\phi}_{A}\right) & 0 \\
		0 & 0 & 1
	\end{bmatrix}
	\notag \\
	&\cdot \begin{bmatrix}
		\cos \left(\bar{\phi}_{E}\right) & 0 & -\sin \left(\bar{\phi}_{E}\right) \\
		0 & 1 & 0 \\
		\sin \left(\bar{\phi}_{E}\right) & 0 & \cos \left(\bar{\phi}_{E}\right)
	\end{bmatrix} \cdot \begin{bmatrix}
		x'+\bar{d} \\ y' \\ z'
	\end{bmatrix}
\end{align}
where $\bar{d}$, $\bar{\phi}_A$, and $\bar{\phi}_E$ are the mean distance, azimuth angle, and elevation angle of a cluster. \textcolor{black}{For instance, when these parameters are set as $\bar{d}=d_n^T$, $\bar{\phi}_E=\phi_{E,n}^T$, and $\bar{\phi}_A=\phi_{A,n}^T$, then scatterers around $C_n^A$ can be generated according to aforementioned assumptions.}


\subsection{Space-Time Evolution of Channel Parameters}
In order to obtain the CIR, space- and time-varying parameters that need to be calculated and updated can be classified into three categories: 1)
EAoDs and AAoDs of rays in the LCS of each LED element, i.e., $\tilde{\psi}_{ij,E,{\rm L}}^T(t)$, $\tilde{\psi}_{ij,A,{\rm L}}^T(t)$, $\tilde{\psi}_{ij,E,{m_n}}^T(t)$, and $\tilde{\psi}_{ij,A,{m_n}}^T(t)$; 2) angles between rays and the normal of effective scatterers (\textcolor{black}{$p$-th PD}), i.e., $\psi_{ij,m_n}^{S,T}(t)$ and $\psi_{m_n}^{S,R}(t)$ (\textcolor{black}{$\psi_{ij,p,{\rm L}}^R(t)$ and $\psi_{p,m_n}^R(t)$)}; 3) transmission distances of rays, i.e., $D_{ij}(t)$, $d_{ij,m_n}^T(t)$, \textcolor{black}{$d_{m_n}^S(t)$}, and $d_{m_n}^R(t)$. Since the parameters are updated according to geometrical relationships, we first update the corresponding coordinate and orientation parameters. Then details of updating these three kinds of parameters will be introduced in the rest of this section.

\subsubsection{Update Time-varying Coordinates and Rx's Orientation}
Considering movements of clusters and Rx, the cartesian coordinates of \textcolor{black}{$S_{m_n}^{A/Z}$} and the optical Rx in GCS are updated, i.e.,
\begin{equation}
	\mathbf{S}_{m_{n}}^{A/Z}(t)=\mathbf{S}_{m_{n}}^{A/Z}(t_0)+v^{C_{n}^{A/Z}} \cdot\ t \\
	\begin{bmatrix}
		\cos \alpha_{E}^{C_{n}^{A/Z}} \cos \alpha_{A}^{C_{n}^{A/Z}} \\
		\cos \alpha_{E}^{C_{n}^{A/Z}} \sin \alpha_{A}^{C_{n}^{A/Z}} \\
		\sin \alpha_{E}^{C_{n}^{A/Z}}
	\end{bmatrix}
\end{equation}
\begin{equation}
	\mathbf{R}_{\rm PD}(t)=\mathbf{R}_{\rm PD}\left(t_{0}\right)+v^{R} \cdot t\left[\begin{array}{c}
		\cos \alpha_{E}^{R} \cos \alpha_{A}^{R} \\
		\cos \alpha_{E}^{R} \sin \alpha_{A}^{R} \\
		\sin \alpha_{E}^{R}
		\end{array}\right]
\end{equation}
where \textcolor{black}{$\mathbf{S}_{m_{n}}^{A/Z}(t_0)$ ($\mathbf{S}_{m_{n}}^{A/Z}(t)$)} and $\mathbf{R}_{\rm PD}(t_0)$ ($\mathbf{R}_{\rm PD}(t)$) are coordinates of \textcolor{black}{$S_{m_n}^{A/Z}$} and the \textcolor{black}{Rx} at the initial time $t_0$ (time~$t$), respectively. Note that \textcolor{black}{$\mathbf{S}_{m_{n}}^{A/Z}(t_0)$} are consistent with $[x,y,z]^{\rm T}$ in (\ref{coord_s}) and $\mathbf{R}_{\rm PD}(t_0)=[D,0,0]^{\rm T}$.

In addition to considering the translational motion, the random 3D rotation of the optical Rx also needs to be taken into account since angle parameters have a great impact on the VLC channel. \textcolor{black}{Taking the normal direction of the top PD in the ADR as the $z_R$ axis and the plane perpendicular to the $z_R$ axis as the $x_Roy_R$ plane, azimuth angles of the $p$-th side PDs ($1 \leq p \leq N_{\rm P D}-1$) can be calculated as $\omega_{\rm PD}^{p}=\frac{2(p-1) \pi}{N_{\rm PD}-1}$, the elevation angles of side PDs are the same and denoted as $\gamma_{\rm PD}^{p}=\pi / 2-\theta_{\rm PD}^{p}$  \cite{Chen_ADR_2018}. Then, the normal vectors of side PDs in the LCS of the ADR can be expressed as $\widetilde{\boldsymbol{n}}_{\rm PD}^{p}=\left[\cos \gamma_{\rm PD}^{p} \cos \omega_{\rm PD}^{p}, \cos \gamma_{\rm PD}^{p} \sin \omega_{\rm PD}^{p}, \sin \gamma_{\rm PD}^{p}\right]^{\mathrm{T}}$. In order to support the ADR as the Rx in the proposed model, the orientation angles of PDs in the LCS $(x_R,y_R,z_R)$ (denoted as ${\rm LCS_{\rm PD}}$) need to be transformed into angles in the GCS of the model. According to the geometrical relationship, the transition matrix $M_{\rm GCS-LCS_{PD}}$ from the GCS to the ${\rm LCS}_{\rm PD}$ can be given by \textcolor{black}{(\ref{transition_M_PD}) shown at the bottom of next page.}
\begin{figure*}[t]
	\centerline{\includegraphics[width=0.8\textwidth]{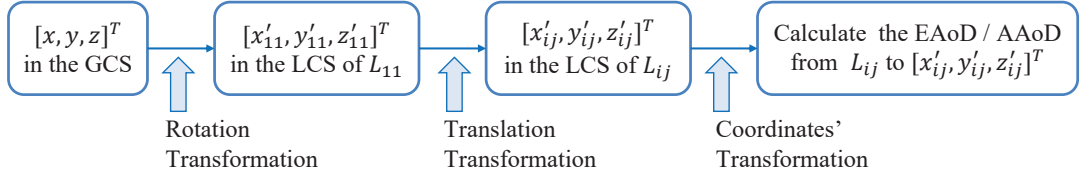}}
	\caption{Flow chart of calculating angles in the LCS of $L_{ij}$.}
	\label{fig_6}
\end{figure*}
\begin{figure*}[b]
	{\noindent} \rule[-10pt]{18.07cm}{0.1em}
	\begin{equation}\label{transition_M_PD}
        	M_{\mathrm{GCS}-\mathrm{LCS}_{\mathrm{PD}}}=\left[\begin{array}{ccc}
        		\cos \beta_{E}^{R} \sin \beta_{A}^{R} & \sin \beta_{E}^{R} \cos \beta_{E}^{R} \cos \beta_{A}^{R} & \cos \beta_{E}^{R} \cos \beta_{A}^{R} \\
        		-\cos \beta_{E}^{R} \cos \beta_{A}^{R} & \sin \beta_{E}^{R} \cos \beta_{E}^{R} \sin \beta_{A}^{R} & \cos \beta_{E}^{R} \sin \beta_{A}^{R} \\
        		0 & -\cos ^{2} \beta_{E}^{R} & \sin \beta_{E}^{R}
        		\end{array}\right]
        \end{equation}
	\begin{equation}\label{transition_M_11}
	M_{{\rm GCS}-{\rm LCS}_{11}}=\begin{bmatrix}
		\cos \beta_{V, E}^{T} \sin \beta_{V, A}^{T} \sin \beta_{H, E}^{T}-\sin \beta_{V, E}^{T} \cos \beta_{H, E}^{T} \sin \beta_{H, A}^{T} & \cos \beta_{V, E}^{T} \cos \beta_{V, A}^{T} & \cos \beta_{H, E}^{T} \cos \beta_{H, A}^{T} \\
		\sin \beta_{V, E}^{T} \cos \beta_{H, E}^{T} \cos \beta_{H, A}^{T}-\cos \beta_{V, E}^{T} \cos \beta_{V, A}^{T} \sin \beta_{H, E}^{T} & \cos \beta_{V, E}^{T} \sin \beta_{V, A}^{T} & \cos \beta_{H, E}^{T} \sin \beta_{H, A}^{T} \\
		\cos \beta_{V, E}^{T} \cos \beta_{H, E}^{T} \sin \left(\beta_{H, A}^{T}-\beta_{V, A}^{T}\right) & \sin \beta_{V, E}^{T} & \sin \beta_{H, E}^{T}
	\end{bmatrix}
        \end{equation}
              \begin{equation}\label{betaA_Cn}
        	\textcolor{black}{\beta_{A}^{C_n^{A/Z}}}={\rm mod} \left[ 2\pi-\arctan \left( \frac{d_n^T\cos(\phi_{E,n}^T)\sin(\phi_{A,n}^T)}{d_{\rm tmp}-d_n^T\cos(\phi_{E,n}^T)\cos(\phi_{A,n}^T)} \right),2\pi \right]
        \end{equation}
        \begin{equation}\label{betaE_Cn}
        	\textcolor{black}{\beta_{E}^{C_n^{A/Z}}}={\rm mod} \left[ -\arctan \left( \frac{d_n^T\sin(\phi_{E,n}^T)}{\sqrt{[d_n^T\cos(\phi_{E,n}^T)]^2+{d_{\rm tmp}}^2-2d_n^T\cos(\phi_{E,n}^T){d_{\rm tmp}}\cos(\phi_{A,n}^T)}} \right),2\pi \right]
        \end{equation}
\end{figure*}
Then, the normal vectors of side PDs in the GCS can be calculated as $\boldsymbol{n}_{\rm PD}^{p}=M_{\mathrm{GCS}-\mathrm{LCS}_{\mathrm{PD}}} \tilde{\boldsymbol{n}}_{\rm PD}^{p}$. Finally, the orientation angles of side PDs in the GCS, i.e., $\beta_{p,A}^R$ and $\beta_{p,E}^R$, can be obtained by the transformation from Cartesian coordinates to spherical coordinates. At time $t$, azimuth and elevation angles of the normal of PDs are calculated as $\beta_{(p,)E}^R(t)=\beta_{(p,)E}^R+\omega_E^R \cdot t$ and $\beta_{(p,)A}^R(t)=\beta_{(p,)A}^R+\omega_A^R \cdot t$, respectively.
}
\subsubsection{Update Angles in the LCS of Each LED Element}
In order to calculate the radiant power in a given direction of a specific ray, the angles of the ray in GCS need to be transformed into angles in LCS of each LED element. 
For the sake of simplicity, we denote the LCS in $L_{11}$ ($L_{ij}$) as  ${\rm LCS}_{11}$ (${\rm LCS}_{ij}$). The transformation process consists of three steps which are shown in Fig.~{\ref{fig_6}}. The first step in this process is to obtain coordinates of \textcolor{black}{$S_{m_n}^A$}/\textcolor{black}{Rx} in ${\rm LCS}_{11}$. According to the geometrical relationship, the transition matrix $M_{{\rm GCS}-{\rm LCS}_{11}}$ from GCS to ${\rm LCS}_{11}$ is given by \textcolor{black}{(\ref{transition_M_11}) shown at the bottom of this page.}
Then, coordinates in ${\rm LCS}_{11}$ are calculated as
\begin{equation}
	[x_{11}^{\prime}, y_{11}^{\prime}, z_{11}^{\prime}]^{\rm T}=M_{{\rm GCS}-{\rm LCS}_{11}}^{-1}[x,y,z]^{\rm T}
\end{equation}
where superscript $\{\cdot\}^{-1}$ means the inverse of the matrix. Next, coordinates in ${\rm LCS}_{ij}$ can be obtained by translation transformation from coordinates in ${\rm LCS}_{11}$, i.e.,
\begin{equation}
	[x_{ij}^{\prime}, y_{ij}^{\prime}, z_{ij}^{\prime}]^{\rm T}=[x_{11}^{\prime}, y_{11}^{\prime}, z_{11}^{\prime}]^{\rm T}- [0,(j-1)\delta_V^T,(i-1)\delta_H^T]^{\rm T}.
\end{equation}
Finally, the EAoD and AAoD of the ray in ${\rm LCS}_{ij}$ is obtained by transforming the cartesian coordinates $[x_{ij}^{\prime},y_{ij}^{\prime},z_{ij}^{\prime}]^{\rm T}$ to spherical coordinates.

By using the transformation process described above, the space- and time-varying angles $\tilde{\psi}_{ij,E,{\rm L}}^T(t)$ and $\tilde{\psi}_{ij,A,{\rm L}}^T(t)$ can be obtained by substituting $\mathbf{R}_{\rm PD}(t)$ into the process. Similarly, by applying the transformation process to \textcolor{black}{$\mathbf{S}_{m_{n}}^A(t)$}, $\tilde{\psi}_{ij,E,{m_n}}^T(t)$ and $\tilde{\psi}_{ij,A,{m_n}}^T(t)$ will be updated at every time instant.

\subsubsection{Update Angles at Scatterers' and Rx' side}
For LoS components, the normalized transmitting vector of the LoS path from $L_{ij}$ is given by 
\begin{equation}\label{dLoS}
	\mathbf{r}_{ij}(t)=\frac{\mathbf{R}_{\rm PD}(t)-\mathbf{L}_{ij}}{\Vert \mathbf{R}_{\rm PD}(t)-\mathbf{L}_{ij} \Vert}.
\end{equation}
Here, the coordinates of $L_{ij}$ in GCS is given as 
\begin{equation}
	\mathbf{L}_{ij}=\begin{bmatrix}
		\tilde{\delta}_V^T\cos(\beta_{V,E}^T)\cos(\beta_{V,A}^T)+\tilde{\delta}_H^T\cos(\beta_{H,E}^T)\cos(\beta_{H,A}^T) \\  
		\tilde{\delta}_V^T\cos(\beta_{V,E}^T)\sin(\beta_{V,A}^T)+\tilde{\delta}_H^T\cos(\beta_{H,E}^T)\sin(\beta_{H,A}^T) \\ 
		\tilde{\delta}_V^T\sin(\beta_{V,E}^T)+\tilde{\delta}_H^T\sin(\beta_{H,E}^T)
	\end{bmatrix}
\end{equation}
where $\tilde{\delta}_V^T=(j-1)\delta_V^T$ and $\tilde{\delta}_H^T=(i-1)\delta_H^T$. Likewise, the normalized transmitting vectors, \textcolor{black}{propagation vector via scatterers}, and receiving vector of the $m$-th ray in the $n$-th path can be determined as
\begin{equation}
	\mathbf{r}_{ij,m_n}^T(t)=\frac{\mathbf{S}_{m_n}^A(t)-\mathbf{L}_{ij}}{\Vert{\mathbf{S}_{m_n}^A(t)-\mathbf{L}_{ij}}\Vert}
\end{equation}
\begin{equation}
	\mathbf{r}_{m_n}^S(t)=\frac{\mathbf{S}_{m_n}^Z(t)-\mathbf{S}_{m_n}^A(t)}{\Vert{\mathbf{S}_{m_n}^Z(t)-\mathbf{S}_{m_n}^A(t)}\Vert}
\end{equation}
\begin{equation}
	\mathbf{r}_{m_n}^R(t)=\frac{\mathbf{R}_{\rm PD}(t)-\mathbf{S}_{m_n}^Z(t)}{\Vert{\mathbf{R}_{\rm PD}(t)-\mathbf{S}_{m_n}^Z(t)}\Vert}
\end{equation}
At scatterers' side, the equivalent normal \textcolor{black}{$\mathbf{n}_{C_n^{A/Z}}$ of $C_n^{A/Z}$} is described with two key parameters, i.e., azimuth angle \textcolor{black}{$\beta_{A}^{C_n^{A/Z}}$ and elevation angle $\beta_{E}^{C_n^{A/Z}}$}. \textcolor{black}{Take the SB case as an example}, the equivalent normal of a cluster is defined as the normalized perpendicular vector of LoS path ($L_{11}$-\textcolor{black}{Rx}) through the center of the cluster\cite{Alsalami2019}.
Depending on geometrical relationships, the azimuth angle and elevation angle of \textcolor{black}{$\mathbf{n}_{C_n^{A/Z}}$} are expressed as \textcolor{black}{(\ref{betaA_Cn}) and (\ref{betaE_Cn}), shown at the bottom of this page}
where $d_{\rm tmp}=d_n^T\cos(\phi_{A,n}^T)\cos(\phi_{E,n}^T)$. On the basis of law of cosines, the AoAs of the ray impinging on the \textcolor{black}{$S_{m_n}^{A}$} can be calculated as
\begin{equation}
	\psi_{ij,m_n}^{S,T}(t)=\arccos \left(-\mathbf{r}_{ij,m_n}^T(t) \cdot \textcolor{black}{\mathbf{n}_{C_n^{A}}} \right)
\end{equation}
while the AoD of the ray reflecting off the \textcolor{black}{$S_{m_n}^{Z}$} is given as
\begin{equation}
	\psi_{m_n}^{S,R}(t)=\arccos \left( \textcolor{black}{\mathbf{n}_{C_n^{Z}}} \cdot \mathbf{r}_{m_n}^R(t) \right). 
\end{equation}
At the Rx's side, the angles between rays and the normal of the \textcolor{black}{$p$-th PD} are determined by 
\begin{equation}
	\psi_{ij,p,{\rm L}}^R(t)=\arccos \left( \mathbf{n}_{\rm PD}^p(t),-\mathbf{r}_{ij}(t) \right)
\end{equation}
\begin{equation}
	\psi_{p,m_n}^R(t)=\arccos \left(\mathbf{n}_{\rm PD}^p(t),-\mathbf{r}_{m_n}^R(t) \right).
\end{equation}
Finally, all of calculations above can be realized as
\begin{align}
	\theta_{\mathbf{X},\mathbf{Y}}&=\arccos [ \cos(\phi_E^X)\cos(\phi_E^Y)\cos(\phi_A^X-\phi_A^Y) \notag \\
	&+\sin(\phi_E^X)\sin(\phi_E^Y) ]
\end{align}
where $\theta_{\mathbf{X},\mathbf{Y}}$ is the angle between vector $\mathbf{X}$ and $\mathbf{Y}$, $\phi_A^X$ ($\phi_A^Y$) and $\phi_E^X$ ($\phi_E^Y$) are azimuth angle and elevation angle of vector $\mathbf{X}$ and $\mathbf{Y}$, respectively.

\subsubsection{Update Propagation Distances of Rays}
The space- and time-varying propagation distances of LoS rays can be calculated as the norm of corresponding vectors, i.e., $D_{ij}(t)=\Vert{\mathbf{R}_{\rm PD}(t)-\mathbf{L}_{ij}}\Vert$. In the same way, the propagation distances of the $m$-th ray in \textcolor{black}{$C_n^{A/Z}$} are given by $d_{ij,m_n}^T(t)=\Vert{\mathbf{S}_{m_n}^A(t)-\mathbf{L}_{ij}}\Vert$, $d_{m_n}^S(t)=\Vert{\mathbf{S}_{m_n}^Z(t)-\mathbf{S}_{m_n}^A(t)}\Vert$, and $d_{m_n}^R(t)=\Vert{\mathbf{R}_{\rm PD}(t)-\mathbf{S}_{m_n}^Z(t)}\Vert$.


\section{\textcolor{black}{Statistical Properties of the Proposed Model}}
\subsection{Channel Transfer Function (CTF)}
The space- and time-varying channel transfer function $H_{ij,p,\lambda_T}(t,f)$ is derived as the Fourier transform of corresponding CIR w.r.t. $\tau$, i.e., 
\begin{equation}\label{CTF}
	H_{ij,p,\lambda_T}(t,f)=\int_{-\infty}^{\infty} h_{ij,p,\lambda_T}(t,\tau) e^{-j 2 \pi f \tau} d \tau.
\end{equation}
By substituting CIR equations of the LoS and NLoS components, the CTF is further written as

\begin{align}\label{CTF_detail}
	H_{ij,p,\lambda_T}(t,f)&=P_{ij,p}^{\rm L}(t)e^{-j2\pi f \tau_{ij}^{\rm L}(t)} \notag \\
	&+ \sum_{n=1}^{N_{ij}(t)}\sum_{m=1}^{M_n} P_{ij,p,\lambda_T,m_n}^{\rm N}(t) e^{-j2\pi f \tau_{ij,m_n}(t)}. 
\end{align}

\begin{figure*}[b]
	{\noindent} \rule[-10pt]{18.07cm}{0.1em}
        \begin{equation}\label{STFCF_L}
        	R_{ij,p,\tilde{i}\tilde{j},\tilde{p},\lambda_T}^{\rm LoS}(t,f;\Delta t,\Delta f) = P_{ij,p,\lambda}^{\rm L}(t) P_{\tilde{i}\tilde{j},\tilde{p},\lambda_T}^{\rm L}(t+\Delta t)e^{j2\pi\left(f\left[\tau_{\tilde{i}\tilde{j},\tilde{p}}^{\rm LoS}(t+\Delta t)-\tau_{ij,p}^{\rm LoS}(t)\right]+\Delta f\tau_{\tilde{i}\tilde{j},\tilde{p}}^{\rm LoS}(t+\Delta t)\right)}
        \end{equation} 
        \begin{align}\label{STFCF_N}
            R_{ij,p,\tilde{i}\tilde{j},\tilde{p},\lambda_T}^{\rm NLoS}(t,f;\Delta t,\Delta f) &= P_{\rm remain}(\delta_i,\delta_j) \times \mathbb{E}\left\{\sum_{n=1}^{N_{ij}(t)}\sum_{n^{\prime}=1}^{N_{\tilde{i}\tilde{j}}(t)}\sum_{m=1}^{M_n}\sum_{m^{\prime}=1}^{M_n}P_{ij,p,m_n,\lambda_T}^{\rm N}(t)P_{\tilde{i}\tilde{j},\tilde{p},m^{\prime}_{n^{\prime}},\lambda_T}^{\rm N}(t+\Delta t) \notag\right. \\
        	\phantom{=\;\;}
        	&\left. \cdot e^{j2\pi\left(f\left[\tau_{\tilde{i}\tilde{j},\tilde{p},m^{\prime}_{n^{\prime}}}^{\rm NLoS}(t+\Delta t)-\tau_{ij,p,m_n}^{\rm NLoS}(t)\right]+\Delta f \tau_{\tilde{i}\tilde{j},\tilde{p},m^{\prime}_{n^{\prime}}}^{\rm NLoS}(t+\Delta t)\right)}\right\}
        \end{align}
\end{figure*}
\subsection{\textcolor{black}{STFCF}}
The STFCF is defined as the correlation between $H_{ij,p,\lambda_T}(t,f)$ and $H_{\tilde{i}\tilde{j},\tilde{p},\lambda}^*(t+\Delta t,f+\Delta f)$ and is expressed as
\begin{align}\label{STFCF}
	&R_{ij,p,\tilde{i}\tilde{j},\tilde{p},\lambda_T}(t,f;\Delta t,\Delta f)\notag \\
	&=\mathbb{E}\left\{H_{ij,p,\lambda_T}(t,f)H_{\tilde{i}\tilde{j},\tilde{p},\lambda_T}^*(t+\Delta t,f+\Delta f)\right\}.
\end{align}
By substituting the equations of CIR into (\ref{STFCF}), the STFCF can be further written as the superposition of correlation functions of the LoS and NLoS components, i.e.,
\begin{align}
	R_{ij,p,\tilde{i}\tilde{j},\tilde{p},\lambda_T}(t,f;\Delta t,\Delta f)&=R_{ij,p,\tilde{i}\tilde{j},\tilde{p},\lambda_T}^{\rm LoS}(t,f;\Delta t,\Delta f) \notag \\
	&+ R_{ij,p,\tilde{i}\tilde{j},\tilde{p},\lambda_T}^{\rm NLoS}(t,f;\Delta t,\Delta f)
\end{align}
where the correlations of LoS and NLoS components can be calculated as \textcolor{black}{(\ref{STFCF_L}) and (\ref{STFCF_N}) shown at the bottom of this page, respectively.}
In (\ref{STFCF_N}), $P_{\rm remain}(\delta_i,\delta_j)$ is the joint probability of a cluster survives from $L_{ij}$ to $L_{\tilde{i}\tilde{j}}$ element.
From STFCF, the temporal ACF, spatial cross correlation function (CCF), and frequency correlation function (FCF) can be easily obtained. For instance, by setting $i=\tilde{i}$, $j=\tilde{j}$, $p=\tilde{p}$, and $\Delta f = 0$, the STFCF is reduced to the temporal ACF, i.e.,
\begin{equation}
	R_{ij,p,\lambda_T}^{\rm ACF}(t,f;\Delta t)=\mathbb{E}\left\{H_{ij,p,\lambda_T}(t,f)H_{ij,p,\lambda_T}^{*}(t+\Delta t,f)\right\}.
\end{equation}
Similarly, the spatial CCF can be obtained by setting $\Delta t=0$ and $\Delta f = 0$, i.e.,
\begin{equation}
	R_{ij,p,\tilde{i}\tilde{j},\tilde{p},\lambda_T}^{\rm CCF}(t,f)=\mathbb{E}\left\{H_{ij,p,\lambda_T}(t,f)H_{\tilde{i}\tilde{j},\tilde{p},\lambda_T}^{*}(t,f)\right\}.
\end{equation}
The FCF can be calculated by setting $i=\tilde{i}$, $j=\tilde{j}$, $p=\tilde{p}$, and $\Delta t = 0$, i.e.,
\begin{equation}
	R_{ij,p,\lambda_T}^{\rm FCF}(t,f;\Delta f)=\mathbb{E}\left\{H_{ij,p,\lambda_T}(t,f)H_{ij,p,\lambda_T}^{*}(t,f+\Delta f)\right\}.
\end{equation}

\subsection{Channel DC Gain and Received Power}
For IM/DD-based VLC systems, the channel is an intensity-in intensity-out channel. The channel DC gain is often used to characterize the optical loss of VLC channel. The space- and time-varying channel DC gain considering actual path loss can be expressed as
\begin{align}\label{DC_Gain}
	H_{ij,p,\lambda_T}(t,0)&=\int_{-\infty}^{\infty}h_{ij,p,\lambda_T}(t,\tau)d\tau \notag \\
	&=P_{ij,p}^{\rm L}(t)+\sum_{n=1}^{N_{ij}(t)}\sum_{m=1}^{M_n} P_{ij,p,\lambda_T,m_n}^{\rm N}(t). 
\end{align}
Since the Rx of a VLC system demodulates the digital signals by detecting the change of received power, the received power is an important channel property. The time-varying received power from $L_{ij}$ is generally defined as
\begin{equation}\label{PR_ij}
	P_{R,ij,p,\lambda_T}(t)=P_{T,ij} \cdot H_{ij,p,\lambda_T}(t,0)
\end{equation}
where $P_{T,ij}$ is the transmitted power from $L_{ij}$. Considering multiple LED lamps in a large LED array, the total received power of \textcolor{black}{$p$-th PD} is calculated as
\begin{equation}\label{PR_total}
	P_{R,p,\lambda_T}(t) = \sum_{i=1}^{M_I}\sum_{j=1}^{M_J}P_{R,ij,p,\lambda_T}(t).
\end{equation}

\subsection{Channel 3dB Bandwidth}
From the perspective of analysis in frequency domain, the channel 3dB bandwidth is obtained as\cite{Al-Kinani2018_survey}
\begin{equation}
	\left|H_{ij,p,\lambda_T}\left(t,f_{3\mathrm{dB}}\right)\right|^{2}=0.5|H_{ij,p,\lambda_T}(t,0)|^{2}.
\end{equation}

\subsection{RMS Delay Spread}
The RMS delay spread is of great significance to characterize the dispersion of propagation delay and is expressed as
\begin{equation}\label{RMS_DS}
\small
	D_{ij,p,\lambda_T,{\rm rms}}(t)=\sqrt{\frac{\int_{-\infty}^{\infty}\left(\tau-\mu_{ij,p,\lambda_T, \tau}(t)\right)^{2} h_{ij,p,\lambda_T}(t,\tau) d \tau}{\int_{-\infty}^{\infty} h_{ij,p,\lambda_T}(t,\tau) d \tau}}
\end{equation}
where the average delay is given as 
\begin{equation}\label{AveD}
	\mu_{ij,p,\lambda_T,\tau}(t)=\frac{\int_{-\infty}^{\infty} \tau \cdot h_{ij,p,\lambda_T}(t, \tau) d \tau}{\int_{-\infty}^{\infty} h_{ij,p,\lambda_T}(t, \tau) d \tau}.
\end{equation}

\subsection{PL}
In VLC channels, large-scale fading due to PL is considered as a prime characteristic. The PL of VLC channel in dB level is given as
\begin{equation}
	{\rm PL}=10\log_{10}(P_{T,{\rm total}}/P_{R,{\rm total}})
\end{equation}
where $P_{T,{\rm total}}$ and $P_{R,{\rm total}}$ are total transmitted and received power of a VLC system at a given distance between Tx and Rx, respectively.

\section{Results and Discussions}
The following section presents results and discussions of key properties of the proposed VLC channel model. Unless otherwise stated, parameters for simulation in this section are set as follows. In the simulation, we consider a scenario where the Rx \textcolor{black}{(single PD)} is pointing to the $L_{11}$ at the initial time. The LED array is a $4\times 4$ array with elevation and azimuth angles setting as $\beta_{H,A}^T~=\pi$, $\beta_{H,E}^T~=\pi/2$, $\beta_{V,A}^T~=\pi/2$, $\beta_{V,E}^T~=0$ and LED spacings setting as $\delta_H^T~=\delta_V^T~=1\ {\rm m}$. The total transmitted power of a LED lamp is set as $P_{\rm Tx}~=1\ {\rm W}$\cite{Zeng2018}. The radiation patterns of LEDs are set as default Lambertian patterns with $\alpha=1$. The parameters of the \textcolor{black}{Rx} are given as $\beta_R^A~=\pi$, $\beta_E^R~=0$, $A_R~=1\ {\rm cm^2}$\cite{Miramirkhani2020}, $\Psi_{\mathrm{FoV}}~=85^{\circ}$\cite{Miramirkhani2020}, and $G(\psi^R)~=T(\psi^R)~=1$. \textcolor{black}{The location parameters of clusters are given as ${\rm std}[\phi_{E,n}^{T(R)}]~={\rm std}[\phi_{A,n}^{T(R)}]~=40^\circ$, $\bar{\phi}_{E,n}^T~=\bar{\phi}_{E,n}^R~=\bar{\phi}_{A,n}^T~=0$, $\bar{\phi}_{A,n}^R~=\pi$. The ratio of SB components in NLoS paths is set as $\eta_{\rm SB}=0.9$.} The number of effective scatterers within a cluster is set as $M_n~=100$\cite{Wang2021}. The effective area of a cluster is set as $A_{c,{\rm eff}}~=1\ {\rm m^2}$, and the effective area of each scatterer is calculated as $A_{m_n,{\rm eff}}=A_{c,{\rm eff}}/M_n$. Considering a typical indoor scenario where Tx is located at the ceiling and Rx is held by the user, the distance from $L_{11}$ to the \textcolor{black}{Rx} at the initial time is set as $D~=2$~m. The effective reflectance parameters of clusters are calculated as (\ref{Gamma}) and randomly generated with a given weight parameter which is described in Appendix~A in detail. The Txs are white LEDs with the wavelength range in $380\ {\rm nm}$\ --\ $780\ {\rm nm}$.


\subsection{STFCF}
 \begin{figure}[tb]
 	\centerline{\includegraphics[width=0.5\textwidth]{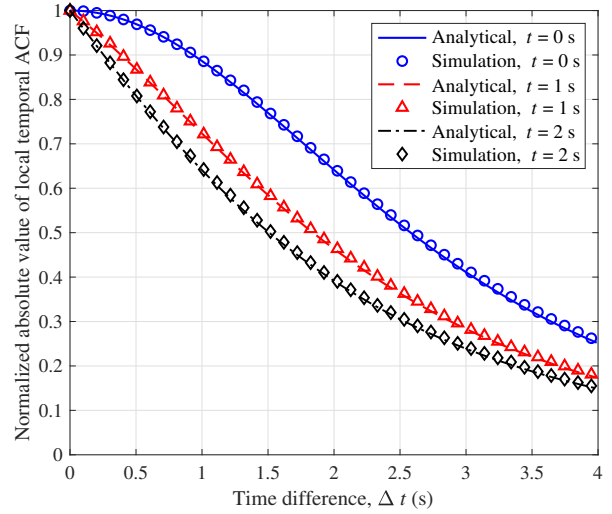}}
 	\caption{Temporal ACFs at different time instants (sub-channel: $L_{11}$-PD, $v^R~=0.5$ m/s, $\alpha_A^R~=0$, $\alpha_E^R~=\pi/2$, $\sigma_{DS}~=\sigma_{AS}~=\sigma_{ES}~=1$ m, $\lambda_B~=80$ /m, $\lambda_D~=4$ /m, $D_c^A=10$ m).}
 	\label{fig_acf}
 \end{figure}
By setting $p=\tilde{p}$ and $q=\tilde{q}$, the \textcolor{black}{STFCF} can be reduced to the temporal ACF, the comparison of temporal ACFs of the sub-channel $L_{11}$-PD at $0$ s, $1$ s, and $2$ s is demonstrated in Fig.~{\ref{fig_acf}}. It can be seen that the temporal ACF is not only related to time differences, but also associated to the time instants. These findings indicate that the channel shows non-stationarity in the time domain resulting from the time-varying channel conditions. Moreover, the analytical results have a good consistency with the simulated results, validating the correctness of the derivations and simulations.

 \begin{figure}[tb]
 	\centerline{\includegraphics[width=0.5\textwidth]{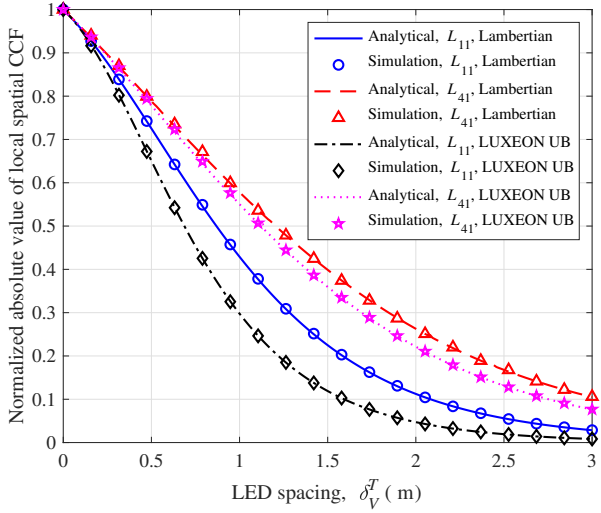}}
 	\caption{Spatial CCFs at different LED elements with different LED radiation patterns ($\sigma_{DS}~=\sigma_{AS}~=\sigma_{ES}~=1$~m, $\lambda_B~=80$ /m, $\lambda_D~=4$ /m, $D_c^A=10$ m, $t~=0$ s).}
 	\label{fig_ccf}
 \end{figure}
Fig.~{\ref{fig_ccf}} shows spatial CCFs of the channel with ideal Lambertian and LUXEON UB radiation patterns at different LED elements. It can be clearly seen that different LED elements correspond to different spatial correlations, showing the non-stationarity of indoor VLC channels in the space domain. In addition, LED radiation patterns can also affect channel spatial CCFs. Since the Lambertian radiation pattern has a larger beamwidth than the LUXEON UB radiation pattern, more multipath components can be observed by adjacent LED elements simultaneously, spatial correlations are enhanced correspondingly. Besides, the consistency between analytical results and simulation results reflects the validity of our simulations and derivations.

\begin{figure}[tb]
	\centerline{\includegraphics[width=0.5\textwidth]{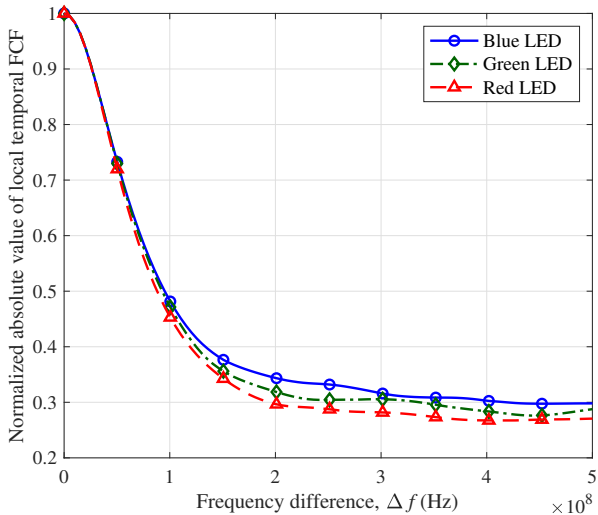}}
	\caption{FCFs with light sources of different colors (sub-channel: $L_{11}$-PD, $\sigma_{DS}~=\sigma_{AS}~=\sigma_{ES}~=1/1.1/1.2$ m for red/green/blue LED).}
	\label{fig_fcf}
\end{figure}

\textcolor{black}{Using light sources of different colors as the Tx, the FCFs of $C_1^A$ are compared in Fig.~{\ref{fig_fcf}}. The differences between FCFs with light sources of different colors (wavelength ranges) show the non-stationarity in the frequency domain.}

\subsection{Channel DC Gain \& Received Power}
\begin{figure}[t]
	\centerline{\includegraphics[width=0.5\textwidth]{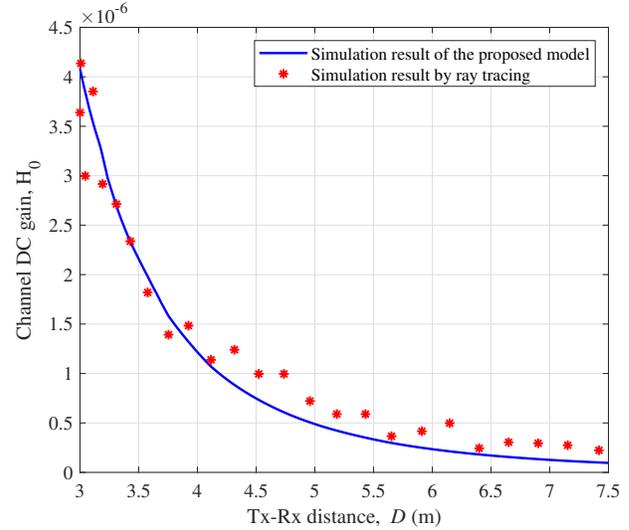}}
	\caption{\textcolor{black}{Channel DC gains of the proposed channel model and the ray tracing results in [30] ($\sigma_{DS}~=\sigma_{AS}~=1.2$~m, $\sigma_{ES}~=1$ m, $N_{c,{\rm total}}=25$, $N_{c,{\rm SB}}=22$, $A_{c,{\rm eff}}=5\ {\rm m^2}$, $v^R=1$ m/s, $\alpha_A^R=0$, $\alpha_E^R=\frac{\pi}{2}$, $D=3$ m, $t=0-7$ s).}}
	\label{fig_fitDCGain}
\end{figure}

\textcolor{black}{Fig.~{\ref{fig_fitDCGain}} presents the comparison of channel DC gains of the proposed model with the ray tracing results in \cite{Miramirkhani2015}. It can be seen that the proposed model can approximate well with ray tracing results in the absence of accurate environmental information, illustrating that the proposed model can obtain good accuracy with lower complexity than the ray tracing model. However, the difference between simulation results due to the approximation error of Lambertian radiation pattern and the real light source radiation pattern used in ray tracing can still be observed.}

\begin{figure}[t]
	\centering
	\subfigure[]{{\includegraphics[width=0.49\textwidth]{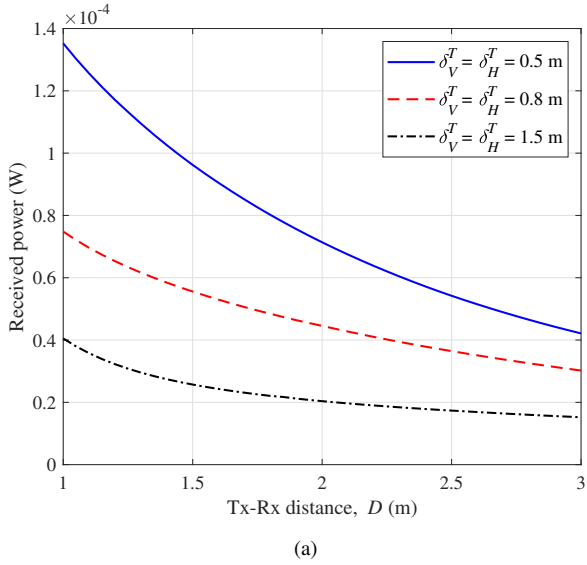}}\label{fig_8}}
	\subfigure[]{{\includegraphics[width=0.49\textwidth]{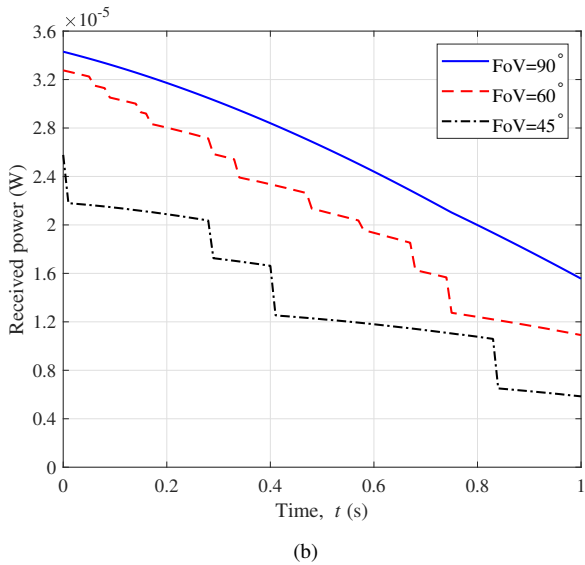}}\label{fig_FoV3}}
	\caption{Total received powers: a) with different distances and different spacings between LED elements ($\Psi_{\rm FoV}=85^{\circ}$, $\sigma_{DS}~=\sigma_{AS}~=\sigma_{ES}~=1$ m, $\lambda_B~=80$ /m, $\lambda_D~=4$ /m, $D_c^A=10$ m, $t~=0$ s), b) \textcolor{black}{when the optical Rx is rotating  with different FoV angles ($\omega_A^R=\pi/4$ rad/s, $\delta_H^T~=\delta_V^T~=1\ {\rm m}$, $\sigma_{DS}~=\sigma_{AS}~=\sigma_{ES}~=1$ m, $\lambda_B~=80$ /m, $\lambda_D~=4$ /m, $D_c^A=10$ m).}}
	\label{fig_RP}
\end{figure}

The relationships between the total received power at the initial time and Tx-Rx distance with different LED elements' spacings are shown in Fig.~{\ref{fig_8}}. The path loss will be higher when the Tx-Rx distance increases, thus the total received power will become lower. The changing trend of the curves is consistent with simulation results in \cite{Al-Kinani2020} and \cite{Al-Kinani2018_V2V}. Meanwhile, it can be seen that the larger the LED elements' spacing, the lower the received power. This is because a larger spacing corresponds to a larger propagation distance, and thus larger path loss.

For the situation where the received power is time-varying, Fig.~{\ref{fig_FoV3}} illustrates simulation results of $P_R(t)$ with the rotation of the Rx. Note that in the simulation, the Rx is pointing to $L_{11}$ directly at the initial time. Therefore, the received power becomes lower with time as the Rx rotates away from the main lobe. What's more, the comparison of time-varying received powers with different field-of-view angles $\Psi_{\rm FoV}$ are presented in Fig.~{\ref{fig_FoV3}}. It can be observed that larger $\Psi_{\rm FoV}$ will enhance the received power due to the fact that the PD will receive more multipath components with larger $\Psi_{\rm FoV}$. Moreover, the received power changes smoothly over time with $\Psi_{\rm FoV}=90^{\circ}$ while it has some abrupt changes with $\Psi_{\rm FoV}=60/45^{\circ}$. This is because that when $\Psi_{\rm FoV}$ is too small, part of multipath components will suddenly disappear in the FoV during the rotation of Rx.

\begin{figure}[t]
	\centering
	\subfigure[Lambertian]{{\includegraphics[width=0.39\columnwidth]{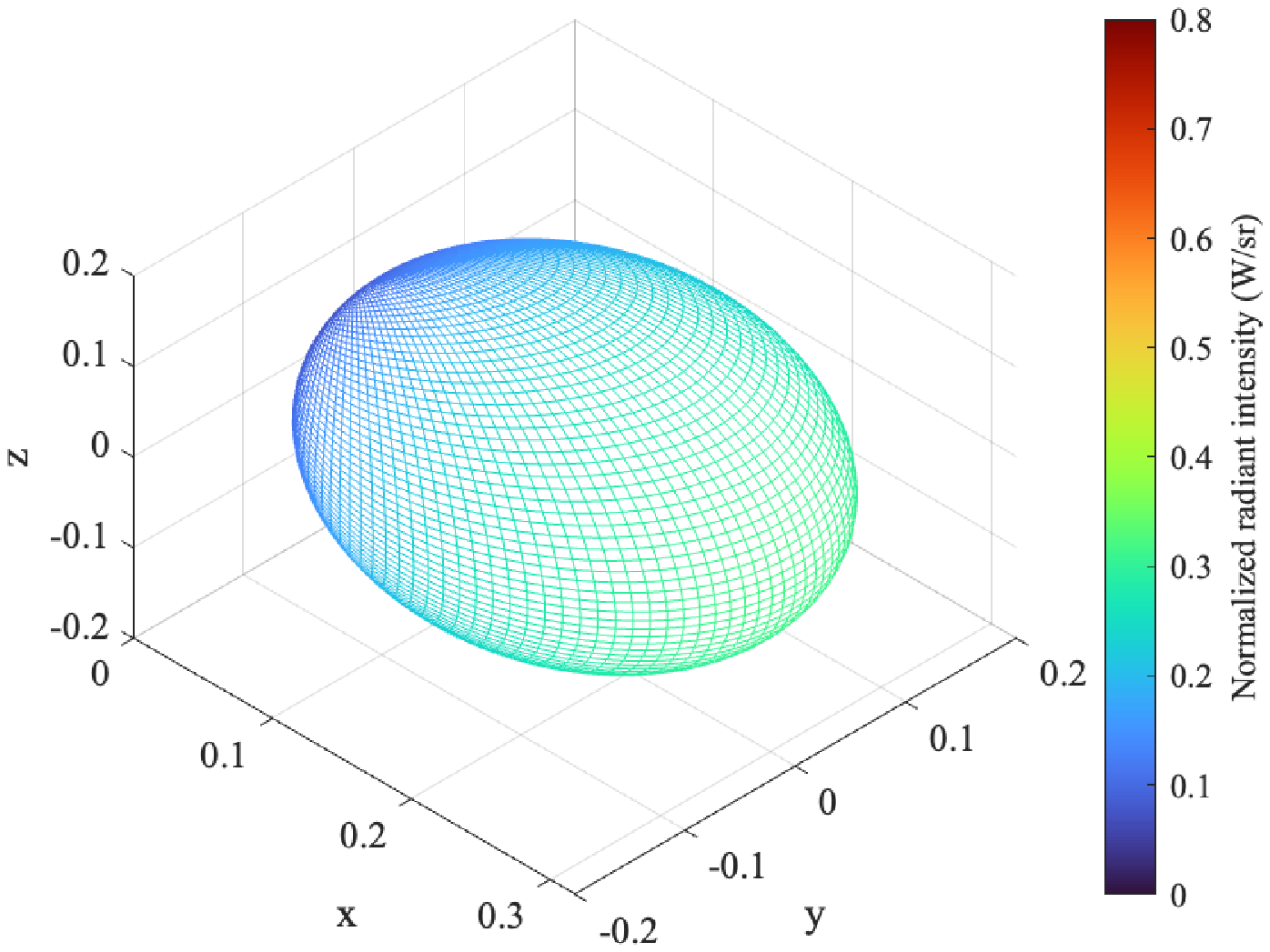}}\label{RadiantP_Lambertian}}
	\subfigure[XLamp]{{\includegraphics[width=0.39\columnwidth]{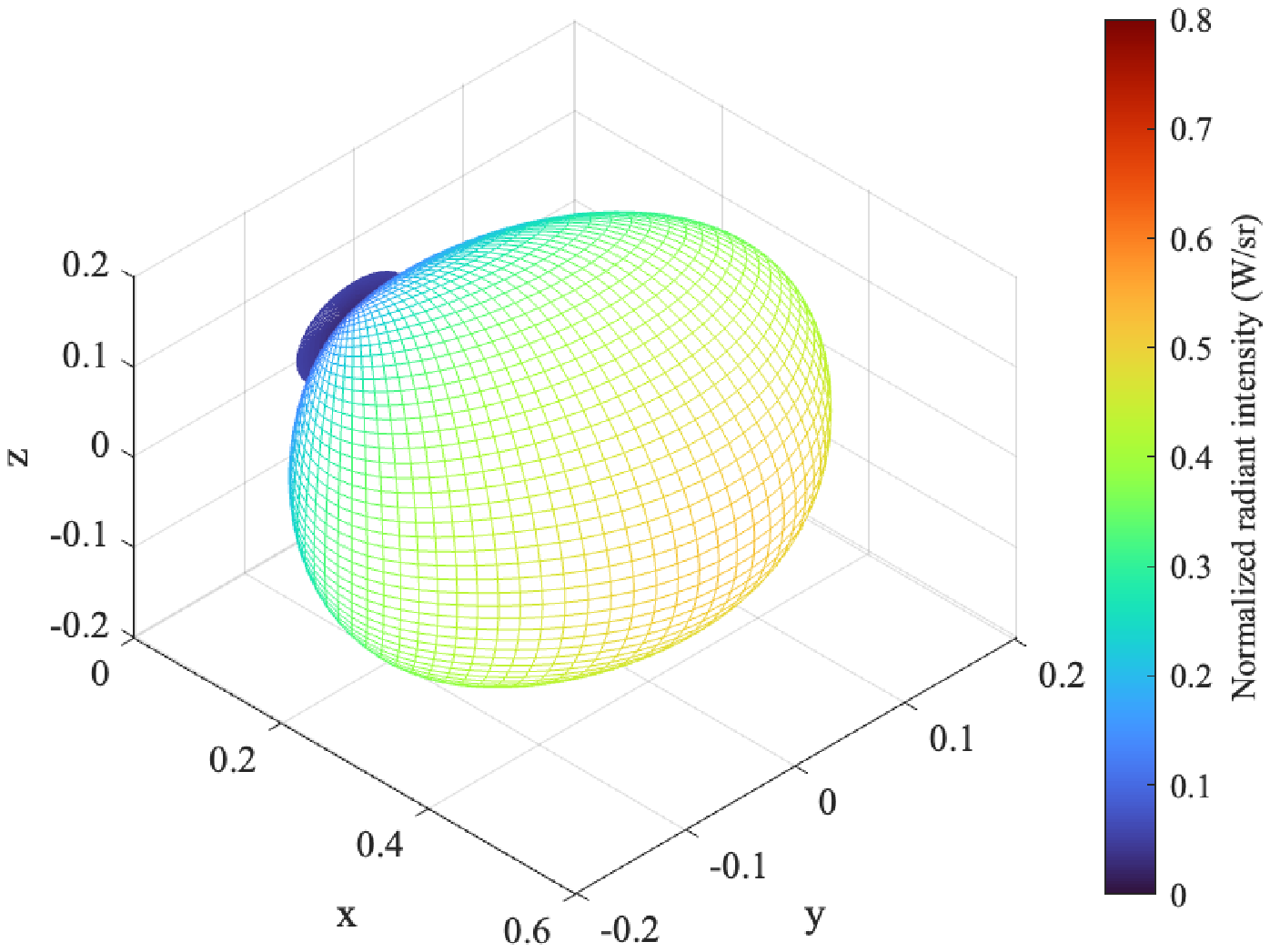}}\label{RadiantP_XLamp}}
	\\
	\subfigure[LUXEON LB]{{\includegraphics[width=0.39\columnwidth]{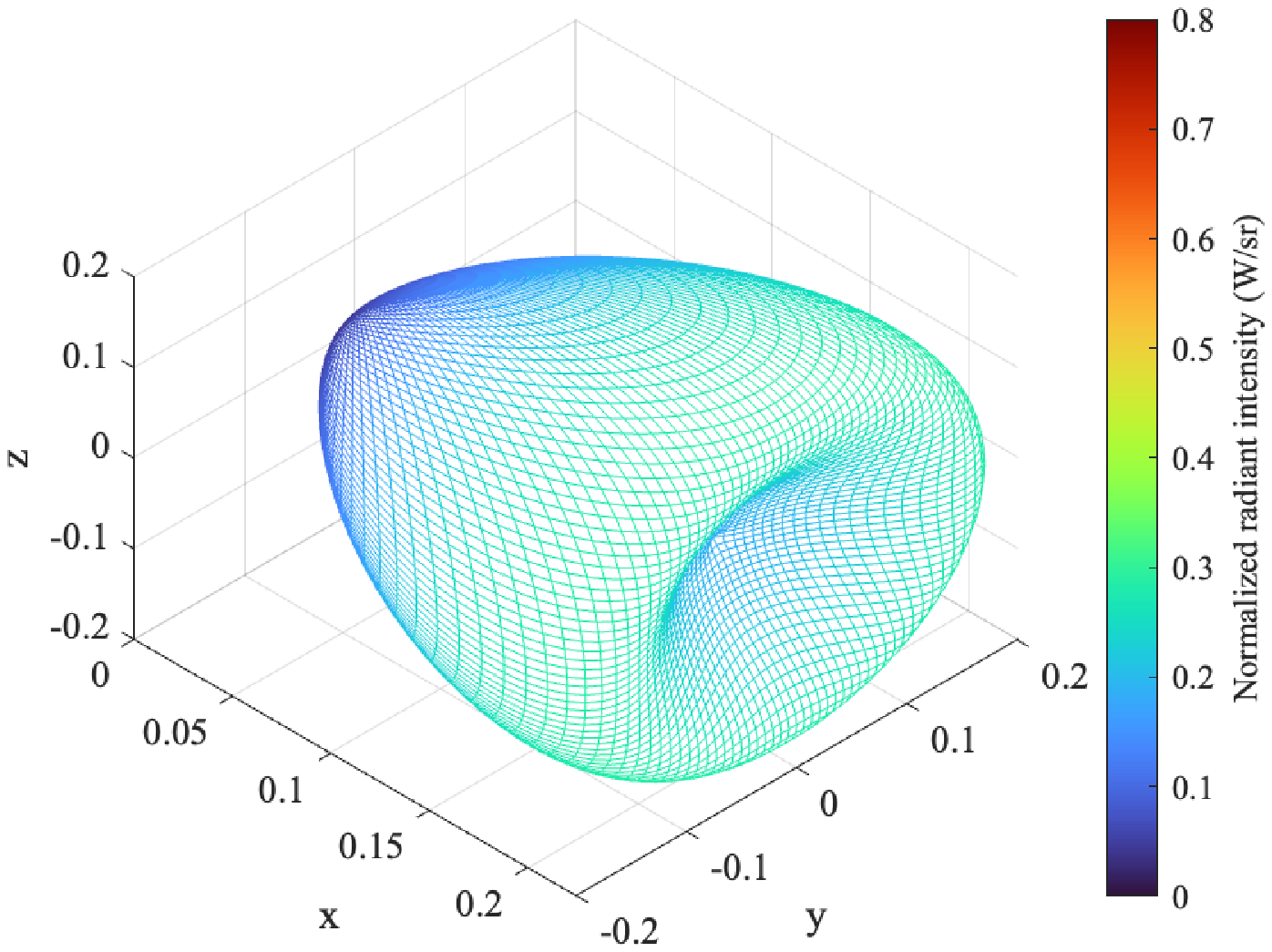}}\label{RadiantP_LUXEON_LB}}
	\subfigure[LUXEON UB]{{\includegraphics[width=0.39\columnwidth]{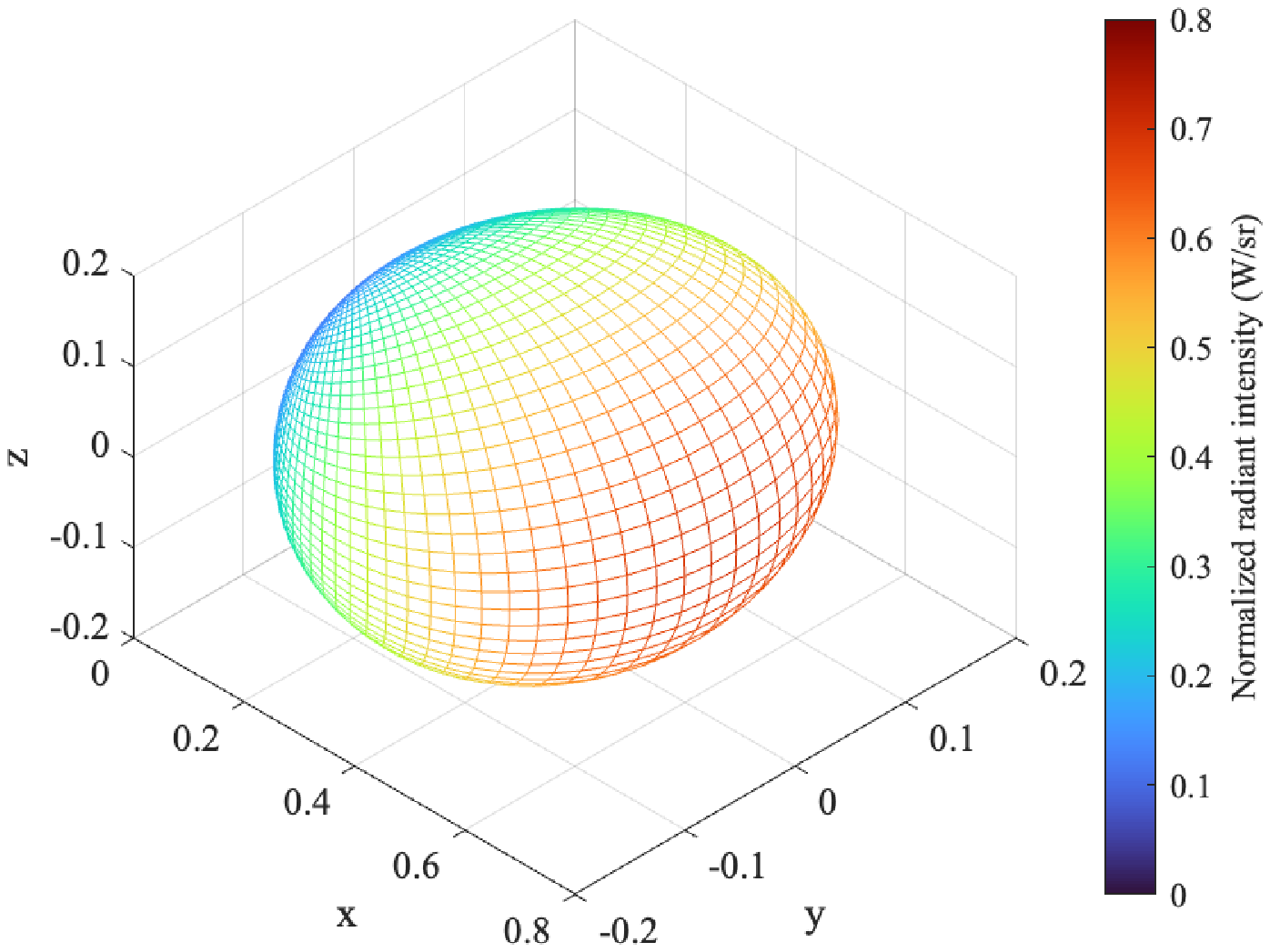}}\label{RadiantP_LUXEON_UB}}
	\caption{\textcolor{black}{3D radiant patterns of four kinds of LEDs.}}
	\label{fig_10}
\end{figure}

\begin{figure}[tb]
	\centerline{\includegraphics[width=0.35\textwidth]{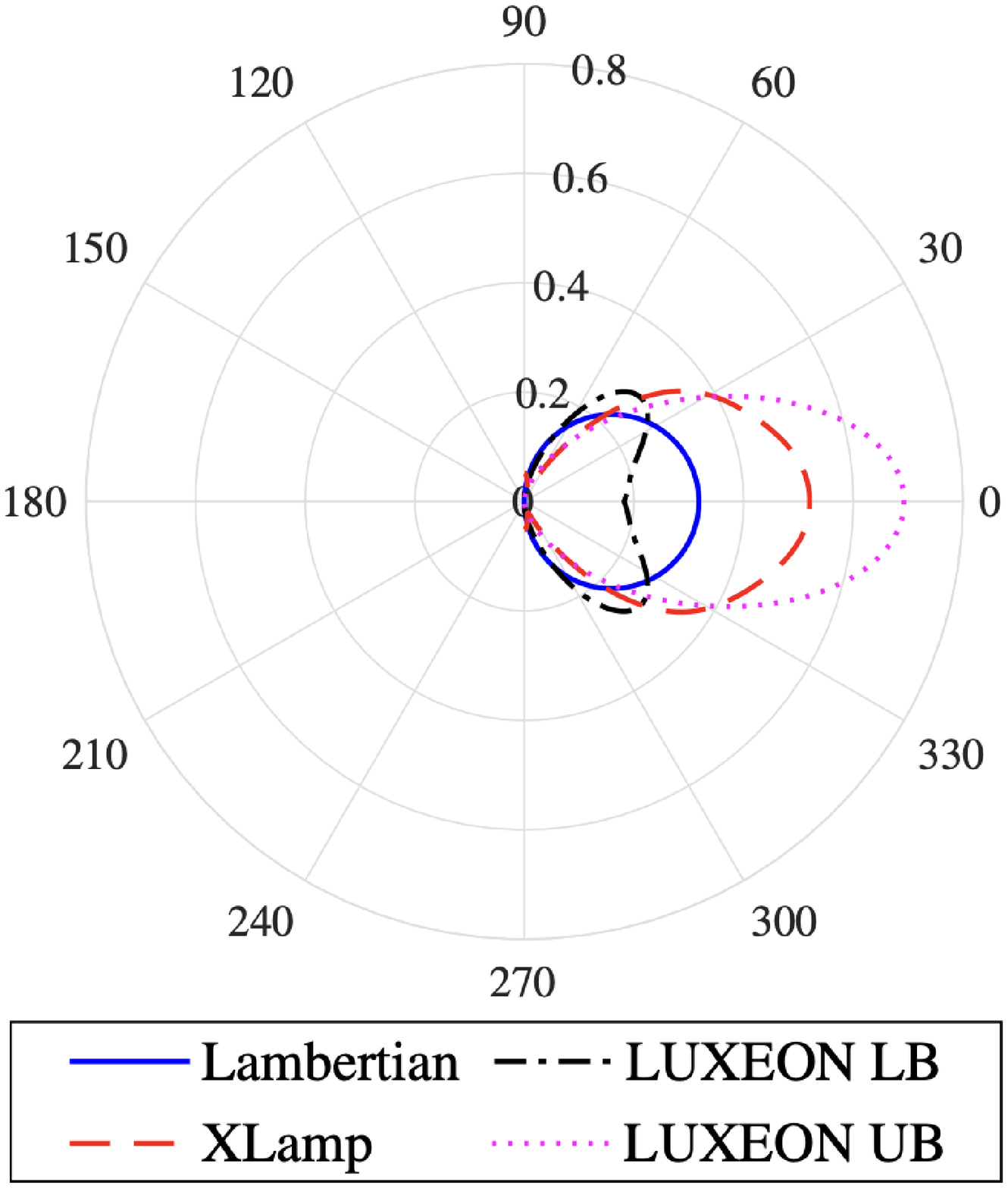}}
	\caption{\textcolor{black}{2D radiant patterns of four kinds of LEDs.}}
	\label{fig_11}
\end{figure}
\subsection{RMS Delay Spread}


\begin{figure}[t]
	\centering
	\subfigure[]{{\includegraphics[width=0.49\textwidth]{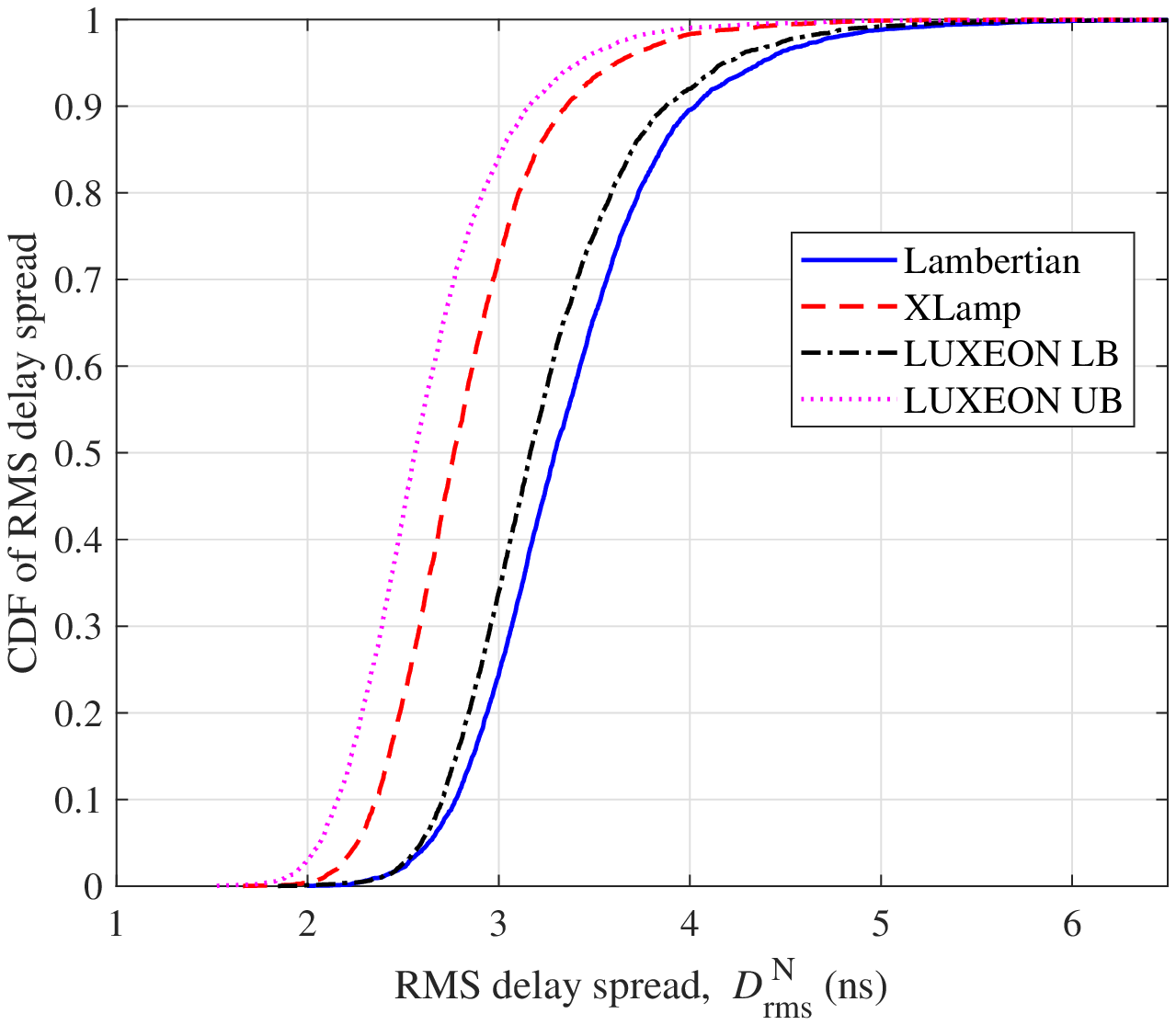}}\label{fig_12a}}
	\subfigure[]{{\includegraphics[width=0.49\textwidth]{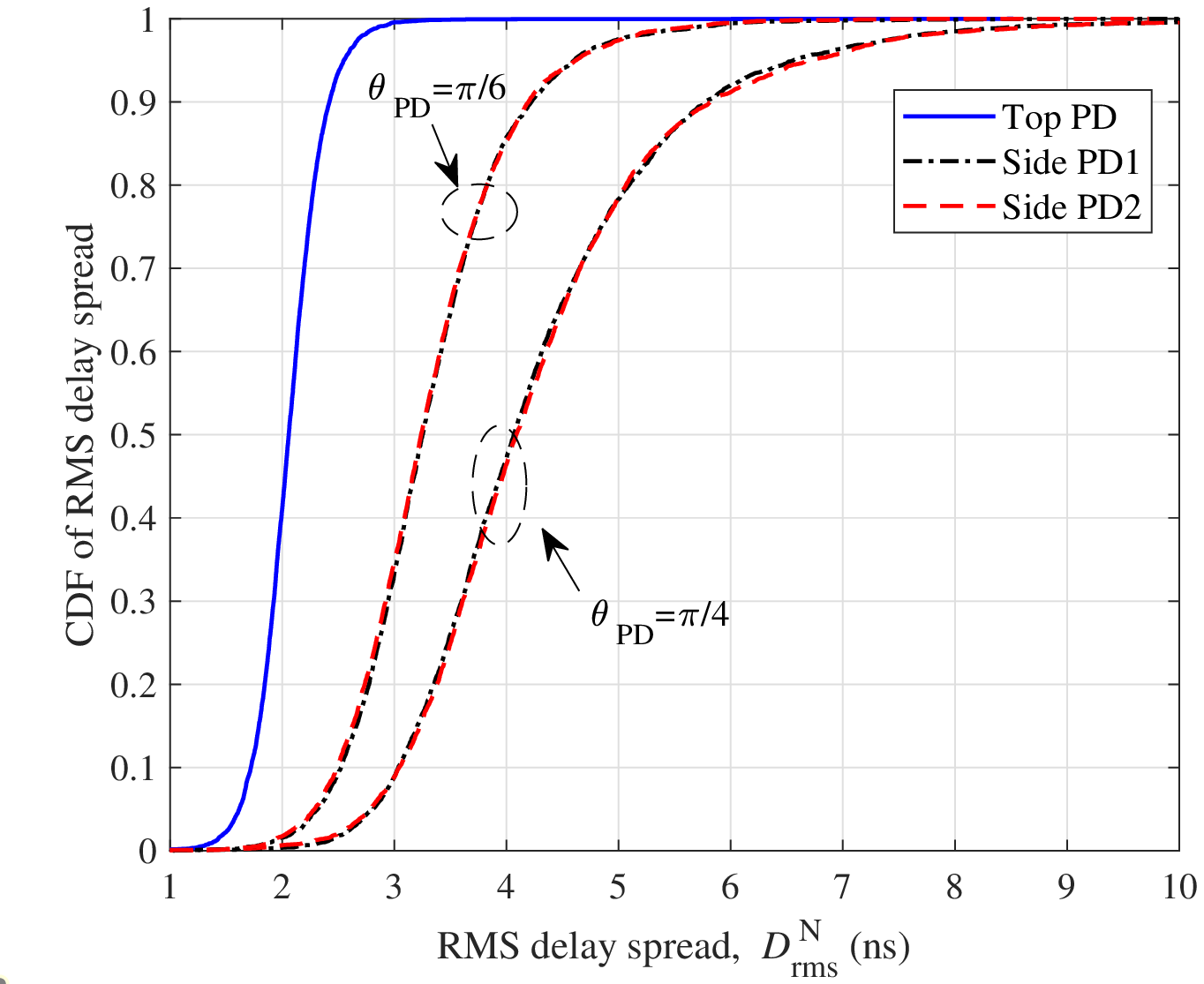}}\label{fig_12b}}
	\caption{CDFs of RMS delay spread: a) of $h_{11,1,\lambda_T}^{\rm N}(t,\tau)$ with different LED radiation patterns (single PD, $\sigma_{DS}~=\sigma_{AS}~=\sigma_{ES}~=1$~m, $\lambda_B~=80$ /m, $\lambda_D~=4$ /m, $D_c^A=10$ m, $t~=0$ s), \textcolor{black}{b) of different sub-channels in the ADR ($N_{\rm PD}=3$, $\Psi_{\mathrm{FoV}}=60^{\circ}$, $\lambda_B~=80$ /m,~$\lambda_D~=4$ /m, $D_c^A=10$ m, $t~=0$ s).}}
	\label{fig_12}
\end{figure}


Unlike most of existing VLC channel models, the proposed channel model can support any special radiation pattern of LED. In this work, we try to simulate and analyze VLC channel properties under settings of four LED radiation patterns as shown in Fig.~{\ref{fig_10}} and Fig.~{\ref{fig_11}}. The ideal Lambertian radiation pattern obtained from (\ref{Lambertian}) is presented in Fig. \ref{RadiantP_Lambertian}. The other three kinds of radiation patterns illustrated in Fig.~{\ref{fig_10}} can be found in \cite{Moreno2008}. Note that the radiation pattern models in \cite{Moreno2008} require corresponding angle transformations and normalizations before substituting into our channel model for simulation. As is shown in Fig.~{\ref{fig_10}}, the four radiation patterns have special shapes and radiation distributions. From another aspect, we can observe different beamwidths of these four radiation patterns in Fig.~{\ref{fig_11}}. Fig.~{\ref{fig_12a}} shows the comparison of cumulative distribution functions (CDFs) of RMS delay spread of $h_{11,1,\lambda_T}^{\rm N}(t,\tau)$ with the four LED radiation patterns. It can be observed that LED radiation patterns will influence the channel delay spread. When the LED lamp have larger beamwidth, it will observe more multipath components, thus making the channel delay spread larger. Due to the fact that ideal Lambertian ($\alpha=1$) and the LUXEON lower bound (LB) (XLamp and LUXEON upper bound (UB)) radiation patterns have similar beamwidths, the differences of corresponding delay spread results are relatively small. Since commercially available LEDs usually have different special radiation patterns, it is necessary to support special patterns in the channel model.

\textcolor{black}{Fig.~{\ref{fig_12b}} illustrates the CDFs of RMS delay spread of different sub-channels in the ADR. An obvious difference between the delay spread of the top PD sub-channel and those of the side PDs can be seen from the curves, while CDFs of delay spread of side PDs in the same ADR are almost the same. The comparison of CDFs reveals that the top PD needs to be separated from side PDs for corresponding processing when designing VLC systems equipped with an ADR. In addition, elevation angles of side PDs in the ADR also have a significant impact on the channel.}



\subsection{PL}

\begin{figure}[t]
	\centering
	\subfigure[]{{\includegraphics[width=0.49\textwidth]{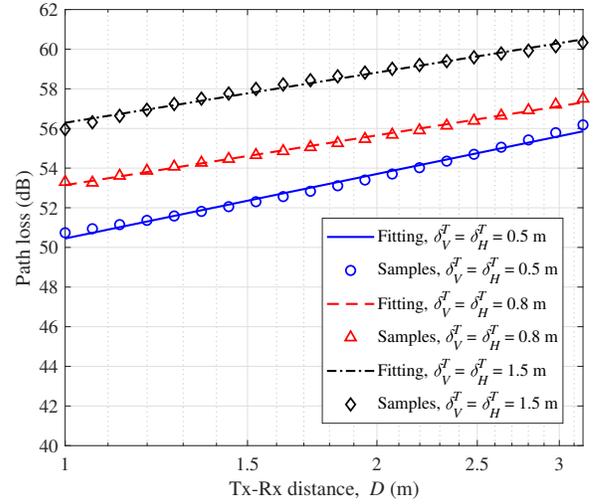}}\label{fig_PL_delta}}
	\subfigure[]{{\includegraphics[width=0.49\textwidth]{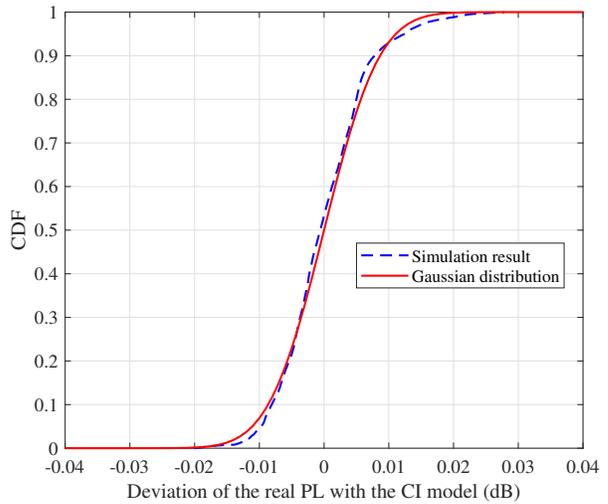}}\label{fig_shadowing}}
	\caption{a) PLs with different spacings between LED elements, b) \textcolor{black}{The CDF of the deviation of the real PL of the sub-channel $L_{11}-$PD with the CI model in dB level ($\sigma_{DS}~=\sigma_{AS}~=\sigma_{ES}~=1$ m, $\lambda_B~=80$ /m, $\lambda_D~=20$ /m, $D_c^A=10$ m, $t~=0$ s).}}
	\label{fig_PL}
\end{figure}

In terms of PL, simulation and fitting results of PLs with different LED elements' spacings are given in Fig. \ref{fig_PL_delta}. The samples are obtained through multiple simulations by changing the distance between $L_{11}$ and the Rx. We try to fit the simulation samples with the CI reference distance PL model $P L(d)=P L(d_0)+10 \gamma \log (d/d_0)$, where $d_0$ is the reference distance, $\gamma$ denotes the propagation coefficient. As can be seen in Fig. \ref{fig_PL_delta}, PLs of indoor VLC channels can fit well with the CI reference distance model. The PL will be larger with larger LED elements' spacing which is consistent with the trend in Fig.~{\ref{fig_8}}.



\textcolor{black}{The CDF of the deviation of the real PL with the CI model (exactly the shadowing) is compared with the Gaussian distribution in Fig.~{\ref{fig_shadowing}}. As can be seen, the statistical property of the shadowing in dB level fits well with the Gaussian distribution, which means the shadowing will show a lognormal distribution in the linear domain. This phenomenon once again shows that the multipath superposition of real-valued signals in VLC channels will cause the large-scale shadowing fading.}

\subsection{Channel 3dB Bandwidth}
\begin{figure}[tb]
	\centerline{\includegraphics[width=0.5\textwidth]{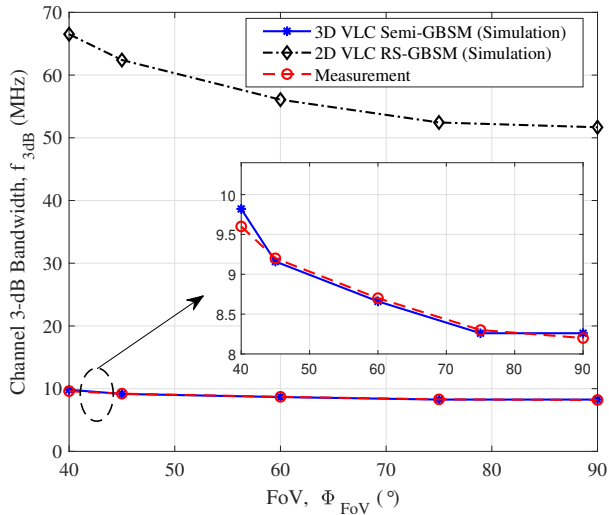}}
	\caption{Channel 3dB bandwidths with different FoVs of the proposed model, the 2D \textcolor{black}{VLC RS-GBSM} in [36], and the measurement data in [51] ($\lambda_1~=\lambda_2~=445\ {\rm nm}$, ${\rm std}[\phi_{E,n}]~=45^\circ$, ${\rm std}[\phi_{A,n}]~=45.5^\circ$, $\bar{\phi}_{E,n}^T~=\pi/12$, $\bar{\phi}_{A,n}^T~=\pi/3$, $N_{c,{\rm total}}~=10$, $M_n~=150$, $\sigma_{DS}~=3.422$ m, $\sigma_{AS}~=2.691$ m, $\sigma_{ES}~=3.719$ m, $D~=2.6345$ m).}
	\label{fig_15}
\end{figure}
In Fig.~{\ref{fig_15}}, the relationship between channel 3dB bandwidth and FoV is simulated and compared with simulation result of the 2D RS-GBSM in \cite{Al-Kinani2016_2} and the measurement data in~\cite{Zhang2012}. The channel measurement was conducted with a blue-light ($445\ {\rm nm}$) single-input single-output (SISO) VLC system in a typical indoor room. Corresponding model parameters are set in accordance with the measurement campaign \cite{Zhang2012} and the rest are chosen according to the estimation procedure introduced in \cite{Wu2018} by fitting with the measurement data in light of the minimum mean square error criterion. It can be clearly seen that the simulation result of the proposed 3D VLC semi-GBSM is in good agreement with the measurement data, showing that our model can well support this communication scenario. \textcolor{black}{Meanwhile, the 2D VLC RS-GBSM in \cite{Al-Kinani2016_2} cannot fit with measurement data well as it oversimplifies the propagation environment and ignores the wavelength-dependency of the VLC channel}. \textcolor{black}{More specifically, the 2D model cannot characterize the influence of angles in 3D environment, and based on the ideal purely geometric assumption which deviates greatly from reality. It can be obviously seen that the proposed model is more accurate and practical.}

\section{Conclusions}
In this paper, a novel 3D \textcolor{black}{space-time-frequency} non-stationary GBSM has been proposed for indoor \textcolor{black}{MIMO} VLC systems. The proposed VLC GBSM can support 3D \textcolor{black}{transitional and rotational} motions of the optical Rx, arbitrary LED radiation patterns, and \textcolor{black}{can be applied to ADRs}. In addition, the space-time evolution of the channel caused by the large LED array and continuous movement of the Rx \textcolor{black}{as well as the wavelength-dependency of light waves have }been taken into consideration. Based on the proposed GBSM, several key statistical properties have been investigated, i.e., \textcolor{black}{STFCF}, channel DC gain, received power, channel 3dB bandwidth, RMS delay spread, and PL. The angle parameters and distance parameters have a great influence on VLC channels. \textcolor{black}{Simulation results have illustrated that our model can mimic the non-stationarities of indoor VLC channels in \textcolor{black}{spatial, time, and frequency} domains.} \textcolor{black}{Moreover, it has been demonstrated that radiation pattern with wider beamwidth corresponds to larger delay spread and higher spatial correlation. The difference in channel characteristics of different PDs in the ADR needs to be considered in the system design. The multipath superposition of real-valued signals in VLC channels will cause large-scale shadowing fadings.} Finally, the fact that the proposed 3D VLC semi-GBSM fits better with measurement data than the existing 2D RS-GBSM has demonstrated the accuracy and practicality of the proposed channel model. In our future work, we will try to extend the proposed channel model to support more VLC scenarios, e.g., outdoor V2V scenario, RIS-aided VLC scenario, etc.

\appendices
\section{\small Calculation of Effective Reflectance $\Gamma_{ij,{\lambda_T}, n}$ in (\ref{Gamma})}
Considering the wavelength-dependent property of VLC channels, an effective reflectance parameter $\Gamma_{ij,\lambda_T,n}$ given in ({\ref{Gamma}}) is introduced and applied into the proposed channel model. To calculate $\Gamma_{ij,\lambda_T,n}$, data of two critical parameters are needed, i.e., the normalized wavelength-dependent radiant PSD of LED $\Phi_{ij}(\lambda)$ and the wavelength-dependent reflectance of clusters $\rho_n(\lambda)$. In \cite{Lee2011,Miramirkhani2020,XLamp}, we can find some data for these parameters. Firstly, we obtain data samples from figures of $\Phi_{ij}(\lambda)$ and $\rho_n(\lambda)$ for several common indoor materials in \cite{Lee2011,Miramirkhani2020,XLamp}. Then the built-in function \emph{trapz} in MATLAB is used to calculate the numerical integration. 
In addition, $\Phi_{ij}(\lambda)$ is normalized to 1 in our calculation to make sure that $\int_{\lambda_1}^{\lambda_2}\Phi_{ij}(\lambda) d\lambda=1$ and $0 \le \Gamma_{ij,\lambda_T,n} \le 1$.

Although data of many common indoor materials are available in \cite{Lee2011} and \cite{Miramirkhani2020}, we consider four kinds of materials that most likely to reflect the visible light in indoor scenarios, i.e., floor, pine wood (furniture), plaster (wall), and plate glass (window). In the simulation, we preset a weight parameter to assign the randomly generated $\Gamma_{ij,\lambda_T,n}$ to different clusters. The weight parameters for these four materials are given as 0.3, 0.2, 0.4, and 0.1 to obtain the results in Section~\uppercase\expandafter{\romannumeral5} except Fig. {\ref{fig_fitDCGain}} (with parameters 0.4, 0.2, 0.4, and 0, respectively).

\bibliographystyle{IEEEtran}

\begin{IEEEbiography}[{\includegraphics[width=1in,height=1.25in,clip,keepaspectratio]{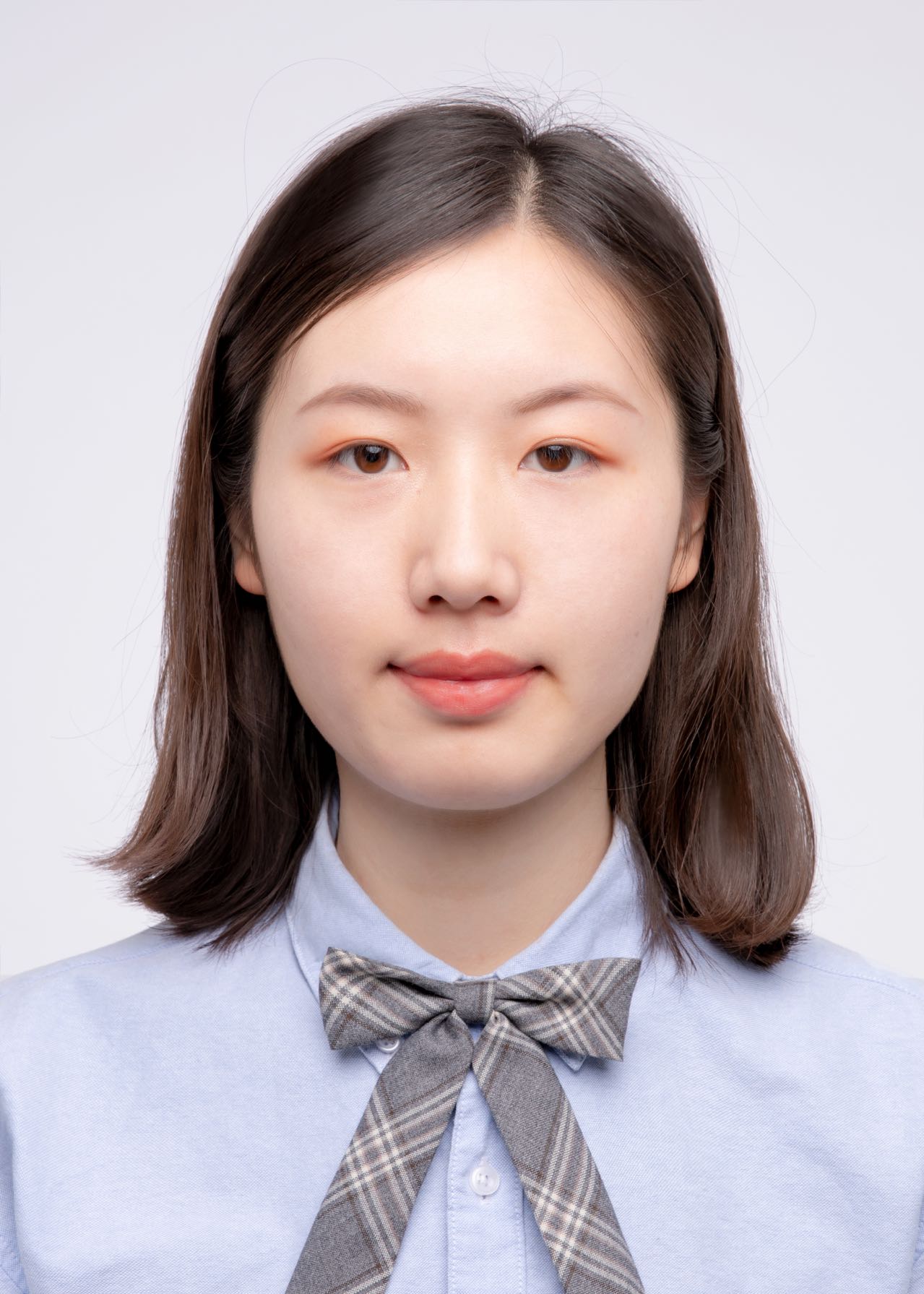}}]{Xiuming Zhu}
	received the B.E. degree in Communication Engineering from Harbin Institute of Technology at Weihai, China, in 2020. She is currently pursuing the M.Sc. degree in the National Mobile
	Communications Research Laboratory, Southeast University, China. Her research interests are optical wireless channel measurements and modeling.
\end{IEEEbiography}

\begin{IEEEbiography}[{\includegraphics[width=1in,height=1.25in,clip,keepaspectratio]{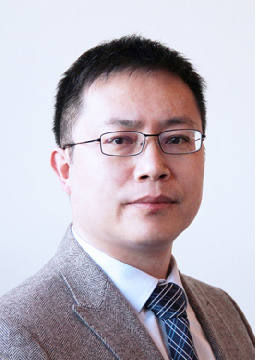}}]{Cheng-Xiang Wang}
(Fellow, IEEE) received the B.Sc. and M.Eng. degrees in communication and information systems from Shandong University, Jinan, China, in 1997 and 2000, respectively, and the Ph.D. degree in wireless communications from Aalborg University, Aalborg, Denmark, in 2004. 

He was a Research Assistant with the Hamburg University of Technology, Hamburg, Germany, from 2000 to 2001, a Visiting Researcher with Siemens AG Mobile Phones, Munich, Germany, in 2004, and a Research Fellow with the University of Agder, Grimstad, Norway, from 2001 to 2005. He has been with Heriot-Watt University, Edinburgh, U.K., since 2005, where he was promoted to a Professor in 2011. In 2018, he joined Southeast University, Nanjing, China, as a Professor. He is also a part-time Professor with Purple Mountain Laboratories, Nanjing. He has authored 4 books, 3 book chapters, and more than 460 papers in refereed journals and conference proceedings, including 25 highly cited papers. He has also delivered 23 invited keynote speeches/talks and 9 tutorials in international conferences. His current research interests include wireless channel measurements and modeling, 6G wireless communication networks, and electromagnetic information theory.

Prof. Wang is a Member of the Academia Europaea (The Academy of Europe), a Fellow of the Royal Society of Edinburgh, IEEE, IET, and China Institute of Communications (CIC), an IEEE Communications Society Distinguished Lecturer in 2019 and 2020, and a Highly-Cited Researcher recognized by Clarivate Analytics in 2017-2020. He is currently an Executive Editorial Committee Member of the IEEE TRANSACTIONS ON WIRELESS COMMUNICATIONS. He has served as an Editor for over ten international journals, including the IEEE TRANSACTIONS ON WIRELESS COMMUNICATIONS, from 2007 to 2009, the IEEE TRANSACTIONS ON VEHICULAR TECHNOLOGY, from 2011 to 2017, and the IEEE TRANSACTIONS ON COMMUNICATIONS, from 2015 to 2017. He was a Guest Editor of the IEEE JOURNAL ON SELECTED AREAS IN COMMUNICATIONS, Special Issue on Vehicular Communications and Networks (Lead Guest Editor), Special Issue on Spectrum and Energy Efﬁcient Design of Wireless Communication Networks, and Special Issue on Airborne Communication Networks. He was also a Guest Editor for the IEEE TRANSACTIONS ON BIG DATA, Special Issue on Wireless Big Data, and is a Guest Editor for the IEEE TRANSACTIONS ON COGNITIVE COMMUNICATIONS AND NETWORKING, Special Issue on Intelligent Resource Management for 5G and Beyond. He has served as a TPC Member, a TPC Chair, and a General Chair for more than 80 international conferences. He received 14 Best Paper Awards from IEEE GLOBECOM 2010, IEEE ICCT 2011, ITST 2012, IEEE VTC 2013Spring, IWCMC 2015, IWCMC 2016, IEEE/CIC ICCC 2016, WPMC 2016, WOCC 2019, IWCMC 2020, WCSP 2020, CSPS2021, and WCSP 2021. Also, he received the 2020--2022 “AI 2000 Most Inﬂuential Scholar Award Honourable Mention” in recognition of his outstanding and vibrant contributions in the ﬁeld of Internet of Things.
\end{IEEEbiography}

\begin{IEEEbiography}[{\includegraphics[width=1in,height=1.25in,clip,keepaspectratio]{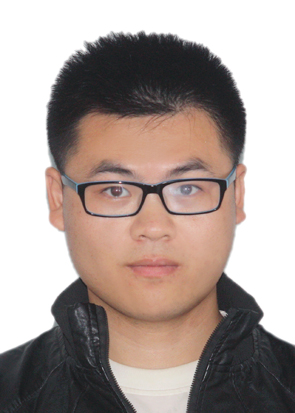}}]{Jie Huang}
	(Member, IEEE) received the B.E. degree in Information Engineering from Xidian University, China, in 2013, and the Ph.D. degree in Communication and Information Systems from Shandong University, China, in 2018. From October 2018 to October 2020, he was a Postdoctoral Research Associate in the National Mobile Communications Research Laboratory, Southeast University, China, supported by the National Postdoctoral Program for Innovative Talents. From January 2019 to February 2020, he was a Postdoctoral Research Associate in Durham University, UK. Since Apr. 2021, he is an Associate Professor in the National Mobile Communications Research Laboratory, Southeast University, China and also a researcher in Purple Mountain Laboratories, China since Mar. 2019. His research interests include millimeter wave, THz, massive MIMO, reconfigurable intelligent surface channel measurements and modeling, wireless big data, and 6G wireless communications. He received 3 Best Paper Awards from WPMC 2016, WCSP 2020, and WCSP 2021. He has also delivered 2 tutorials in IEEE/CIC ICCC 2021 and IEEE PIMRC 2021.
\end{IEEEbiography}

\begin{IEEEbiography}[{\includegraphics[width=1in,height=1.25in,clip,keepaspectratio]{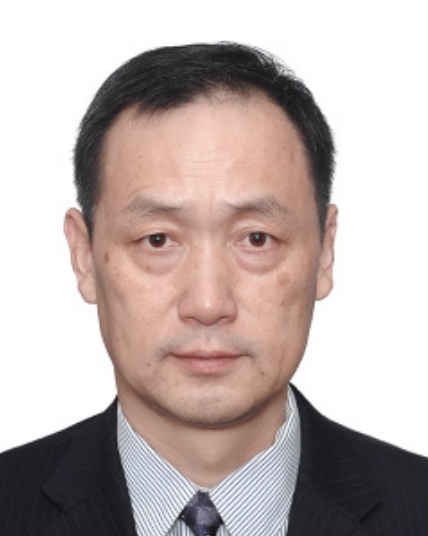}}]{Ming Chen}
	(Member, IEEE) received the B.Sc., M.Sc., and Ph.D. degrees in mathematics from Nanjing University, Nanjing, China, in 1990, 1993, and 1996, respectively. In July 1996, he joined the National Mobile Communications Research Laboratory, Southeast University, as a Lecturer. From April 1998 to March 2003, he was an Associate Professor and has been a Professor with the laboratory since April 2003. His research interests include signal processing and radio resource management of mobile communication systems.
\end{IEEEbiography}

\begin{IEEEbiography}[{\includegraphics[width=1in,height=1.25in,clip,keepaspectratio]{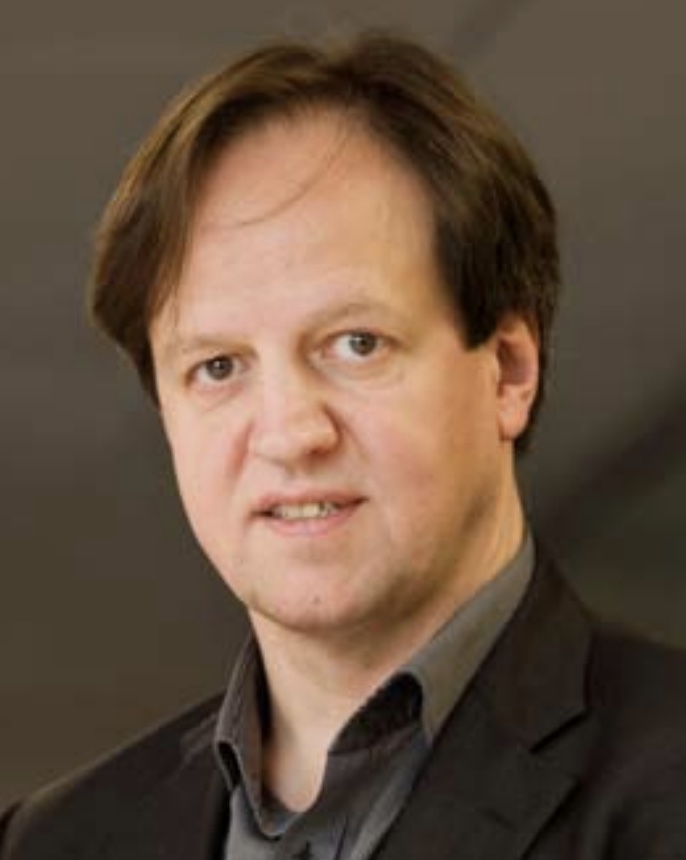}}]{Harald Haas}
	received the Ph.D. degree from The University of Edinburgh in 2001. He is a Distinguished Professor of Mobile Communications at The University of Strathclyde/Glasgow, Visiting Professor at the University of Edinburgh and the Director of the LiFi Research and Development Centre. Prof. Haas set up and co-founded pureLiFi. He currently is the Chief Scientific Officer. He has co-authored more than 600 conference and journal papers. He has been amongthe Clarivate/Web of Science highly cited researchers between 2017-2021. Haas’ main research interests are in optical wireless communications and spatial modulation which he first introduced in 2006. In 2016, he received the Outstanding Achievement Award from the International Solid State Lighting Alliance. He was the recipient of IEEE Vehicular Society James Evans Avant Garde Award in 2019. In 2017, he received a Royal Society Wolfson Research Merit Award. He was the recipient of the Enginuity The Connect Places Innovation Award in 2021. He is a Fellow of the IEEE, the Royal Academy of Engineering (RAEng), the Royal Society of Edinburgh (RSE) as well as the Institution of Engineering and Technology (IET).
\end{IEEEbiography}

\end{document}